\documentclass[11pt,a4paper,twocolumn]{IEEEtran}
\usepackage{epsfig,times,url}
\title{Towards Reliable Network Wide Broadcast in Mobile Ad Hoc
Networks}
\author{Paul Rogers and Nael Abu-Ghazaleh\thanks{
The authors are at the Computer Science Dept., State University of New
York at Binghamton: emails progers1@binghamton.edu, nael@cs.binghamton.edu}}
\begin{document}
\maketitle

\begin{abstract}
Network-Wide Broadcast (NWB) is a common operation in Mobile Ad hoc
Networks (MANETs) used by routing protocols to discover routes and in
group communication operations.  NWB is commonly performed via {\em
flooding}, which has been shown to be expensive in dense MANETs
because of its high redundancy.  Several efforts have targeted
reducing the redundancy of floods.  In this work, we target another
problem that can substantially impact the success of NWBs: since MAC
level broadcasts are unreliable, it is possible for critical
rebroadcasts to be lost, leading to a significant drop in the node
coverage.  This is especially true under heavy load and in sparse
topologies.  We show that the techniques that target reducing the
overhead of flooding, reduce its inherent redundancy and harm its
reliability.  In addition, we show that static approaches are more
vulnerable to this problem.  We then present a selective rebroadcast
approach to improve the robustness of NWBs.  We show that our approach
leads to considerable improvement in NWB coverage relative to a
recently proposed solution to this problem, with a small increase in
overhead.  The proposed approaches do not require proactive neighbor
discovery and are therefore resilient to mobility.  Finally, the
solution can be added to virtually all NWB approaches to improve their
reliability.
\end{abstract}

\section{Introduction}
The network topology in Mobile Ad hoc Networks (MANETs) is defined by
the physical location of the nodes, which changes due to mobility.
Network-Wide Broadcast (NWB) provides a mechanism to deliver
information to nodes in a way that is resilient to topology changes.
Therefore, NWB is a heavily used primitive at the core of most MANET
routing~\cite{clausen-03,das-00} and group communication
protocols~\cite{lewis-02,royer-99a}.  A common approach to performing
NWB is flooding: a process where every node that receives a packet for
the first time, rebroadcasts it.  Flooding has been shown to be
wasteful, especially in dense networks, a problem called {\em the
broadcast storm}~\cite{ni-99}.  Ni et al, who identified the problem,
also proposed several solutions to it based on nodes locally
determining whether their rebroadcast is likely to be needed.  An
alternative, topology-sensitive, approach to the problem attempts to
construct a virtual backbone that is tasked with disseminating the
broadcast.  For example, Connected Dominating Sets (CDS) can be
constructed comprising of a connected subset of the nodes that
together cover all the nodes in the network; this set can then be
tasked with rebroadcasting the NWB packet while all other nodes just
receive it.

In this paper, we consider a different problem that adversely affects
most NWB protocols: the NWB unreliability problem.  This problem
affects NWB protocols that rely on MAC level broadcast operations.
This includes most existing NWB algorithms including flooding based
protocols~\cite{ni-99} and virtual backbone approaches
(e.g.,~\cite{clausen-03,alzoubi-02,ghandi-03}).  More specifically,
because MAC broadcasts are unreliable, it is possible for rebroadcasts
to be lost due to interference or transmission errors.  The loss rate can be 
considerable if high interference exists or if link quality is bad as has been
observed in wireless testbeds~\cite{decouto-02}.  This 
may lead
to the NWB reaching only a subset of the nodes.  A particularly bad
example of this problem occurs when the initial transmission of the
flood packet is lost at all receivers (e.g., due to a collision with
another packet).  In this case the NWB will not reach any nodes.

NWB algorithms that control redundancy to reduce
overhead have increased vulnerability to this problem; redundancy
provides some protection against losses.  This is
especially true for virtual backbone NWB algorithms that statically
determine the set of forwarding nodes: if a transmission to one of
these nodes is lost, the NWB is lost to the remainder of the backbone and
the nodes they cover.  Lou and Wei identified the vulnerability of
these approaches and proposed a solution for addressing it (Double
Covered Broadcast, or DCB)~\cite{lou-04}.  DCB works by constructing
virtual backbone graphs that provide double coverage of all nodes --
every node in the graph is in range of two different nodes in the CDS.
Therefore, two retransmissions would need to be lost before a node is
not covered.  They evaluated this approach and found that it improves
the reliability of CDS approaches without increasing the overhead
substantially.  However, we show that static CDS based approaches
perform worse than dynamic/adaptive approaches in terms of coverage in
lossy environments.

In this paper, we first classify NWB protocols and outline their
properties from a reliability perspective. We show that under high
load (or high error rates) coverage for existing NWB algorithms,
including flooding and DCB, is poor.  We propose a {\em selective
additional rebroadcast} solution to counter potential broadcast
losses.  We show that this solution leads to a substantial improvement
in NWB reliability (much higher than that obtained by DCB and
flooding).  This is especially true in sparse networks where regions
of the network may not have available redundancy for flooding or DCB
to utilize.  The broadcast redundancy is introduced dynamically based
on the observed behavior of the network.  In addition, redundancy
through retransmissions can improve coverage in areas where no
topological redundancy exists, for example, to bridge a critical hop
that connects two partitions of a network.  Because the rebroadcast
decision is carried out locally, based on the observed behavior of the
network, the approach can used with virtually all NWB approaches to
increase their robustness.  Therefore, a combination of an effective
redundancy reduction algorithm, along with selective rebroadcast can
yield a low-overhead algorithm without sacrificing robustness against
interference and transmission errors.

It is important to note that guaranteed reliability requires
acknowledgment from all recipients, which in general, cannot be
accomplished in approaches that rely on MAC broadcast.  For example,
fully reliable NWB can be built on top of unicast communication, but
at a much higher overhead than broadcast based
approaches~\cite{pagani-97,lipman-04}. We do not pursue guaranteed reliability;
rather, we seek to increase the robustness of NWB to MAC broadcast
losses.  Throughout this paper, we use reliability in this context and
not to indicate guaranteed reliability.

The remainder of this paper is organized as follows.
Section~\ref{background} overviews MAC level broadcast and the
difficulties in making them reliable.  Section~\ref{related} presents
an overview of NWB approaches and their properties and reviews other
related work.  Section~\ref{problem} overviews the NWB unreliability
problem and shows its effect on node coverage.  This section also
presents the NWB protocols we use in our studies: we select
representatives of the important classes of NWB algorithms.
Section~\ref{solutions} outlines possible solutions to the
unreliability problem and discusses solutions to it, including
Selective Rebroadcast.  In Section~\ref{experimental} we evaluate
selective rebroadcast of packets using simulation, and show that it
considerably improves NWB reliability for all algorithms.  Coverage
improvement is excellent, but comes at an increase in the NWB
overhead.  Finally Section~\ref{conclude} presents some concluding
remarks.

\section{Background -- MAC Broadcast}~\label{background}
Wireless communication properties make wireless transmissions
unreliable for two primary reasons: (1) Shadowing and obstacles can
cause deep fades in signal power (20dB, or 100x changes in signal
power are common).  This leads to a large number of transmission
errors, especially when the two communicating nodes are far from each
other; and (2) Collisions: the wireless channel is non-uniformly
shared, giving rise to the well known hidden terminal
problems~\cite{bhagarvan-94}.  A hidden terminal is an interfering
node out of reception range of the sender, but in interference range
with the receiver; such a transmission is not detected by the sender
(it is hidden to it), causing a potential collision at the receiver.
Furthermore, techniques such as Carrier Sense Multiple Access and
Collision Avoidance reduce collisions but do not eliminate
them~\cite{fullmer-97}.  

To improve reliability in the face of losses, MAC protocols like IEEE
802.11~\cite{crow-97} use retransmission of packets that are not
acknowledged.  Such an approach cannot be used with broadcast packets
because there are a number of receivers.  Accordingly, if a broadcast
packet is lost due to a collision, {\bf the loss is not detected by
  the sender and no retransmission is carried out}.  This makes MAC
level broadcast operations much more susceptible to losses.  As a
result, NWB operations which rely on MAC level broadcast suffer from
loss of coverage: {\em the NWB unreliability problem}.

\section{Related Work}~\label{related}
In this section, we overview NWB approaches.  We identify the
following three properties: (1) Overhead; (2) resilience to mobility;
and (3) robustness to the NWB unreliability problem.

\subsection{Flooding Based Approaches}
Flooding is the basic approach to NWB: each node rebroadcasts an NWB
packet the first time it receives it.  Flooding is a brute force
approach that has high overhead, especially in dense
networks~\cite{ni-99}.  Most NWB algorithms target this problem.
Flooding does not require topology knowledge; therefore, it is
resilient to mobility and does not require an on-going overhead to
discover neighbors.  Flooding is thought to be resilient to MAC
losses due to the high redundancy generally available in
MANETs. However, we note that in low density networks, or low density
areas of networks, this redundancy is low or even non-existent.

One approach to reducing flooding overhead is to have nodes determine
locally whether their rebroadcast is likely to be redundant.  Ni et
al~\cite{ni-99} suggest several such approaches, where a node decides
not to rebroadcast a packet: (1) probabilistically; (2) based on the
number of rebroadcasts already heard (if several were heard, an
additional one is probably not needed); or (3) based on the distance
or location of the nearest heard rebroadcast (if its too close, it is
likely that little additional coverage is obtained from another
broadcast).  These approaches reduce the overhead without
appreciably harming coverage.  Like flooding, they are resilient to
mobility.  Because each node locally determines whether its
rebroadcast is likely to be needed, the approach dynamically adapts to
transmission losses.

\subsection{Topology Sensitive Approaches} 
An alternative approach is the use of partial network topology
information to build a virtual backbone that can cover all the nodes
in the network.  Only the nodes in the backbone are tasked with
forwarding NWB packets.  Such approaches have been proposed to replace
flooding in routing algorithms~\cite{clausen-03,sivakumar-03}.  A
popular approach to constructing these backbones are Connected
Dominating Sets (CDS) algorithms: a CDS is a subset of the nodes that
is fully connected and sufficient to cover all the nodes in the
network.  While finding the optimal CDS is NP-complete, algorithms
have been developed to construct them distributedly, and
near-optimally in terms of the size of the CDS and the overhead to
construct it (e.g.,~\cite{alzoubi-02,ghandi-03}).

In order to build a virtual backbone, nodes exchange information,
typically about their immediate or two hop neighbors.  Thus, ongoing
overhead is required to discover neighbors.  As a result, this
approach degrades with increasing mobility since the neighborhood
information frequently becomes stale.  Typically the backbone
information is used to statically determine forwarding
responsibilities.  As a result, this approach becomes especially
vulnerable to losses:  if a loss of a packet to a node with forwarding
responsibilities occurs, the remainder of the backbone reachable
through it and the nodes they cover will not receive the NWB.

Another class of topology-sensitive approaches tracks neighbor
  information but dynamically determines membership in the forwarding
  group.  For example, in Flooding with pruning~\cite{lim-00}, a node
  tracks its one hop neighbors and includes them in its broadcast
  retransmission.  Other nodes rebroadcast only if their one hop
  neighbors have not been covered by previous rebroadcasts; this is a
  local decision.  If a particular broadcast is lost, other nodes will
  rebroadcast to compensate since their neighbors are not covered --
  (suboptimal) CDS' are constructed dynamically, making this approach
  more resistant to MAC losses.  However, the overhead is higher, and
  the size of the backbone is likely to be worse, than the static CDS
  approaches because decisions are based on incomplete one-hop
  information that are exchanged with every rebroadcast.  The Scalable
  Broadcast Algorithm~\cite{peng-00} is slightly different: instead of
  including the neighbor-list in the rebroadcast packet, each node
  tracks all its two hop neighbors.  This enables a node A to
  determine which of its neighbors a node B covered by checking the
  two hop neighbor information without requiring B to transmit its
  neighbor list.

In summary, NWB approaches assess forwarding responsibilities either
in a topology sensitive way or based on local estimates of the
rebroadcast importance.  Topology sensitive protocols can provide more
optimal NWBs from an overhead perspective, but require on-going
overhead to exchange neighborhood information and are more susceptible
to mobility.  Within the topology sensitive model, another important
classification is whether the forwarding responsibilities are assigned
statically or dynamically at each forwarding node.  Static
responsibilities allow more optimized forwarding, but are more
susceptible to loss of coverage due to transmission losses.

\subsection{Addressing NWB Unreliability}
All of the approaches above focus primarily on reducing the {\em
overhead} of NWB relative to flooding and therefore, to varying
degrees, result in reducing reliability.  Reliability may be helped
due to the reduction of the contention among the rebroadcast packets,
especially if multiple NWBs are in progress in the network.  For
example, Ghandi et al. attempt to schedule rebroadcasts within a CDS
approach to eliminate self-contention~\cite{ghandi-03}.  Pagani et
al~\cite{pagani-97} and Lipman et al~\cite{lipman-04} propose a fully
reliable unicast-based NWB algorithm.  Unicast based approaches have a
much higher overhead than broadcast-based approaches.

Tang and Gerla proposed modifications to MAC level broadcast to
increase its reliability~\cite{tang-00}.  More specifically, they
introduce acknowledgments to broadcasts: if one or more
acknowledgments (detected as noise on the channel due to collision of
the acknowledgments generated from multiple nodes) are received, it is
assumed that the broadcast is successful; otherwise it is rebroadcast.
Thus, this approach guarantees that at least one node receives a
rebroadcast.  This increases the overhead of broadcast operations
significantly (each receiver must now generate an acknowledgment for
every MAC broadcast).  In addition, because it requires MAC level
modifications, it has a high deployment barrier.

Most similar to our work, Lou and Wei recently identified the effect
of transmission losses on CDS based approaches~\cite{lou-04}.  To counter 
that effect, they proposed a
modified CDS algorithm, called Double Coverage Broadcast (DCB), that
ensures that every node is covered twice (not just once as per the CDS
requirement).  While DCB is more resilient to the NWB unreliability
problem than basic CDS algorithms, the set of forwarding nodes is
statically determined, and therefore, the approach remains vulnerable
to losses.  

\section{Characterizing NWB Unreliability}~\label{problem}
In this section, we characterize the effect of the NWB unreliability
problem, and show that existing NWB algorithms are vulnerable to it.
NWB coverage is affected primarily by two factors: the density of the
network and the probability of MAC transmission loss.  A dense network
has multiple redundant paths that the NWB can follow allowing it to
tolerate some losses without losing too much coverage.  This available
redundancy goes down in sparse networks, or for algorithms that
aggressively control redundancy making these protocols more
unreliabile.  Both interference and transmission losses do not apply
uniformly across the network.  More specifically, interference results
in losses in areas closest to the interfering traffic, while
transmission errors are affected by the surroundings and increase with
the distance between the sender and the receiver. 

The primary performance metrics we are interested in are: (1) Node
coverage: number of nodes that receive the flood; (2) Overhead: number
of retransmissions.  However, since the number of retransmissions is a
function of the number of covered nodes, raw overhead is not
meaningful.  Therefore, we use {\em normalized overhead}, defined as
the number of retransmissions per receiving node, as a measure of
overhead.  For flooding, normalized overhead is always one: every
covered node retransmits the packet once.  Optimized NWB have targeted
lowering the overhead while maintaining coverage which results in a
normalized overhead smaller than 1.

The Network Simulator NS-2 was used for all experiments~\cite{ns-2}.
To generate interference, we used interfering Constant Bit Rate
connections (with different numbers and rates).  In addition, we used
controlled scenarios where losses are initiated by probabilistically
dropping rebroadcast packets at receivers. In all scenarios, nodes are
randomly deployed in a fixed area of 1000 by 1000 meters.  We used the
802.11 MAC implementation in NS-2 with the default parameters (range
of 250 meters).  For CBR interference, the sources and destinations of
the interfering connections were picked randomly.  In each experiment,
one NWB is generated per node.  We time the NWBs to ensure that
successive operations do not interfere.  Each data point represents an
average of 20 scenarios with different random seeds.  To mitigate
unintended routing artifacts from interfering connections (flood
overheads and silent times due to artificial disconnections that may
arise under high load~\cite{xu-02}) we use static routes in the
interfering connections.

In addition to flooding, we also study the following NWB
algorithms:\footnote{We are thankful for Tracy Camp's group at
Colorado Mines for providing us with the base code for LBA, SBA and
AHBP protocols.  We also thank W. Lou for the DCB code.}
\begin{itemize}
\setlength{\itemsep}{0in}
\item Location Based Algorithm (LBA)~\cite{ni-99}: In this algorithm,
  a node includes its location in the rebroadcast packet.  Receiving
  nodes keep track of the rebroadcasts that they overhear.  Based on
  this information a node decides whether its rebroadcast provides
  sufficient coverage to be worth sending.  This is a dynamic
  localized approach that is not topology sensitive.

\item Ad Hoc Broadcast Protocol (AHBP)~\cite{peng-02}: Nodes collect
  two hop neighbor information and use this information to explicitly
  select a set of 1-hop neighbors to rebroadcast the packet such that
  all the 2-hop neighbors are covered.  AHBP is a CDS based approach
  that does not make a local decision on rebroadcasting; instead this
  decision is carried out by the upstream node from whom it received
  the packet.

\item Scalable Broadcast Algorithm (SBA)~\cite{peng-00}: Each node
  maintains a list of two hop neighbors using periodic hello messages.
  When a node A receives a broadcast from node B, it knows the set of
  B's neighbors (from its collected 2-hop neighborhood information).
  If it has additional neighbors that are not covered by B, the node
  schedules a rebroadcast.  This is a dynamic topology sensitive
  algorithm (nodes decide locally whether their rebroadcast is
  necessary given the topology information and previous rebroadcasts).

\item Double-Covered Broadcast (DCB)~\cite{lou-04}: This is a static
  topology-sensitive CDS-based algorithm with built in redundancy
  (double-coverage for every node).
\end{itemize}

\begin{figure}[ht]
\begin{minipage}{3.5in}
\centerline{\epsfig{file=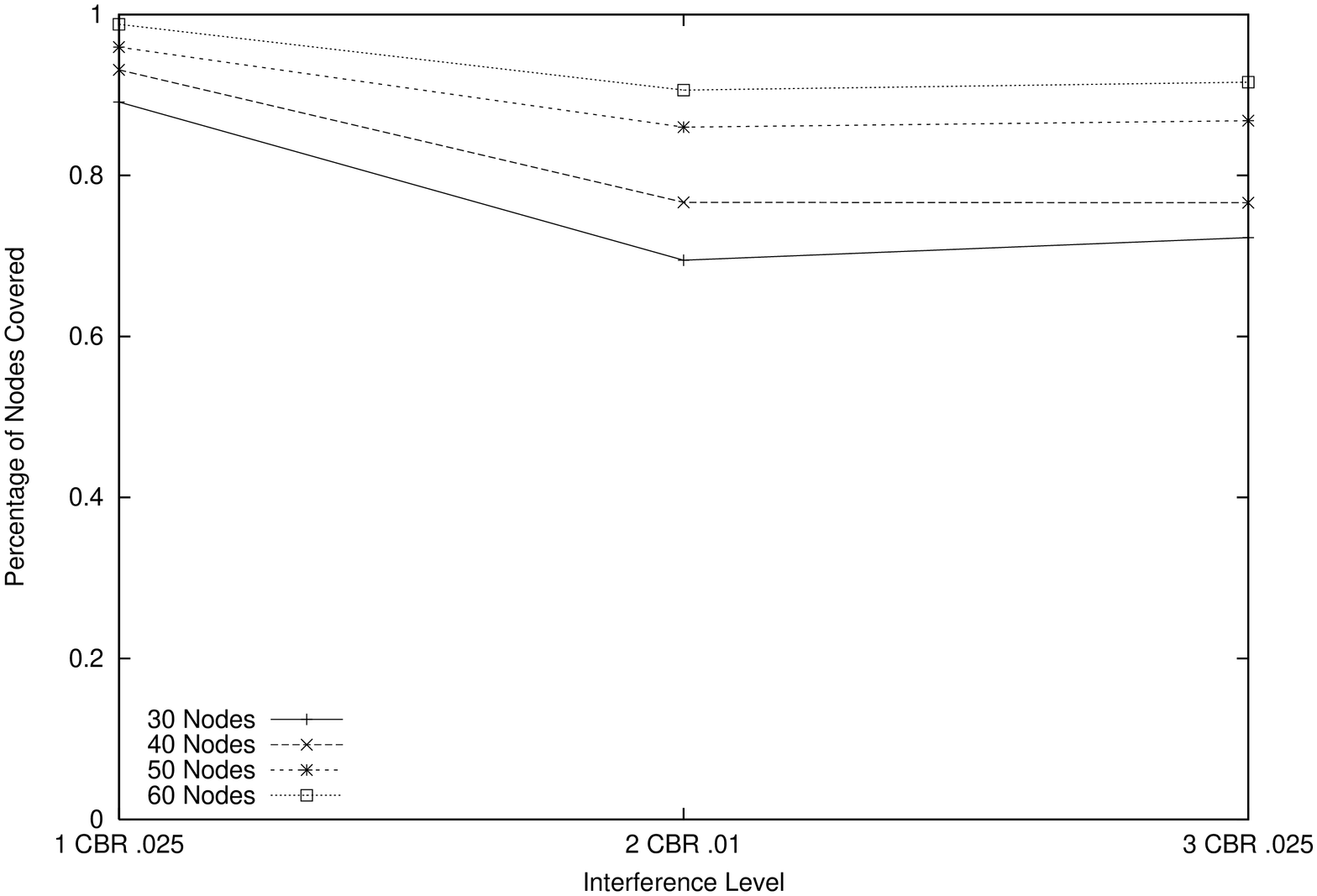,angle=270,width=0.95\linewidth,silent=}}
\caption{{\small Coverage -- CBR Interference}}\label{prob1}
\end{minipage}
\begin{minipage}{3.5in}
\centerline{\epsfig{file=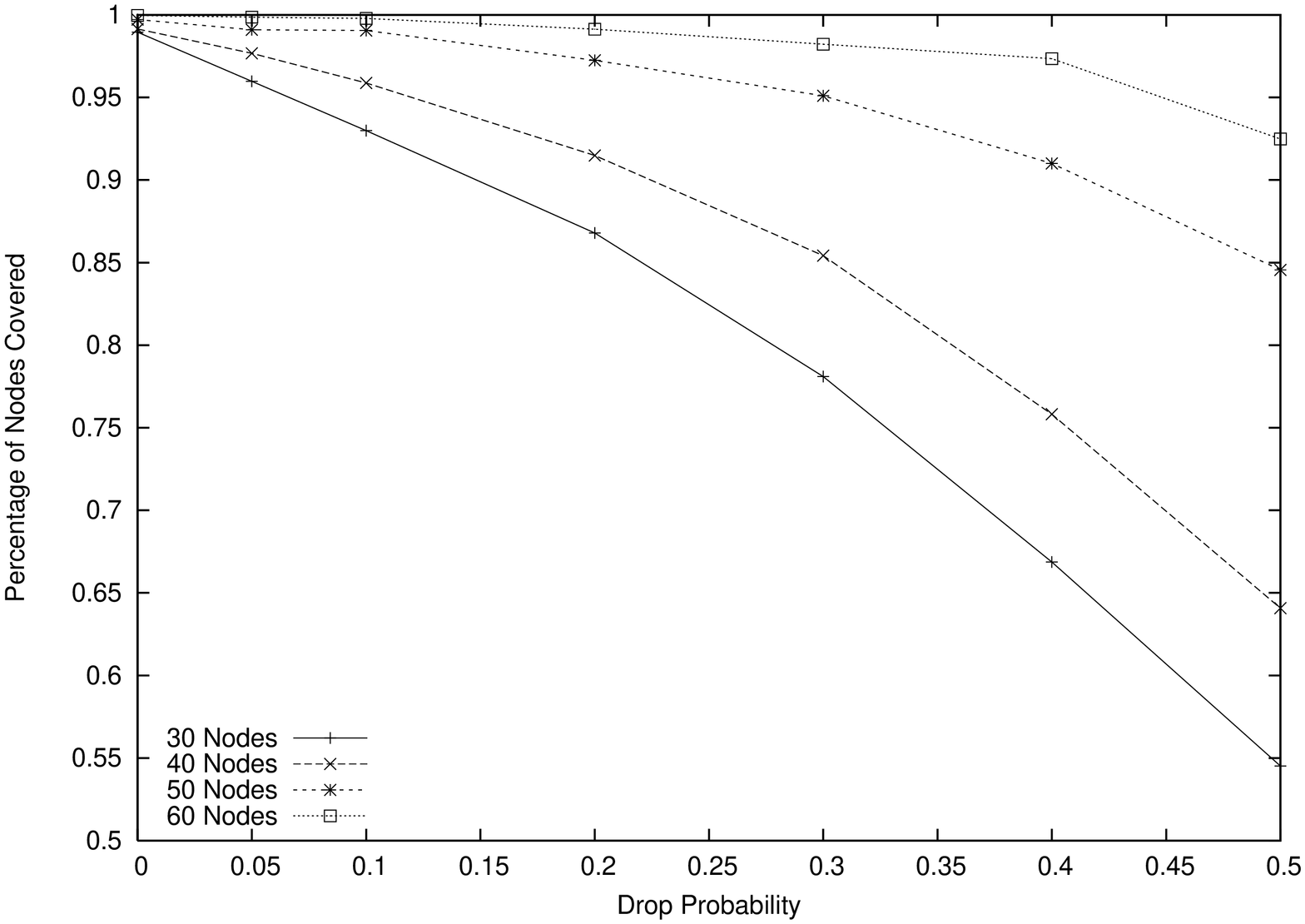,angle=270,width=0.95\linewidth,silent=}}
\caption{{\small  Coverage -- Controlled Drop}}\label{prob2}
\end{minipage}
\end{figure}

Figure~\ref{prob1} and Figure~\ref{prob2} show the node coverage
obtained with flooding for scenarios with randomly deployed nodes.  In
Figure~\ref{prob1} interference is created using competing CBR flows
whose characteristics are shown on the x-axis.  In this case, the
level of interference created by the interfering connections is not
controllable; we do not control the location or the number of hops of
the interfering connection(s).  In contrast, Figure~\ref{prob2}
simulates interference by dropping broadcast packets at the receiver
with a fixed probability.  In this case, it is clear that a
large drop in the node coverage is seen due to interference.
The primary observation is that the severity of the problem is more
pronounced in sparse networks because of the limited redundancy
available in them.  As the density increases, flooding becomes more
resilient to losses.

\begin{figure}[ht]
\begin{minipage}{3.5in}
\centerline{\epsfig{file=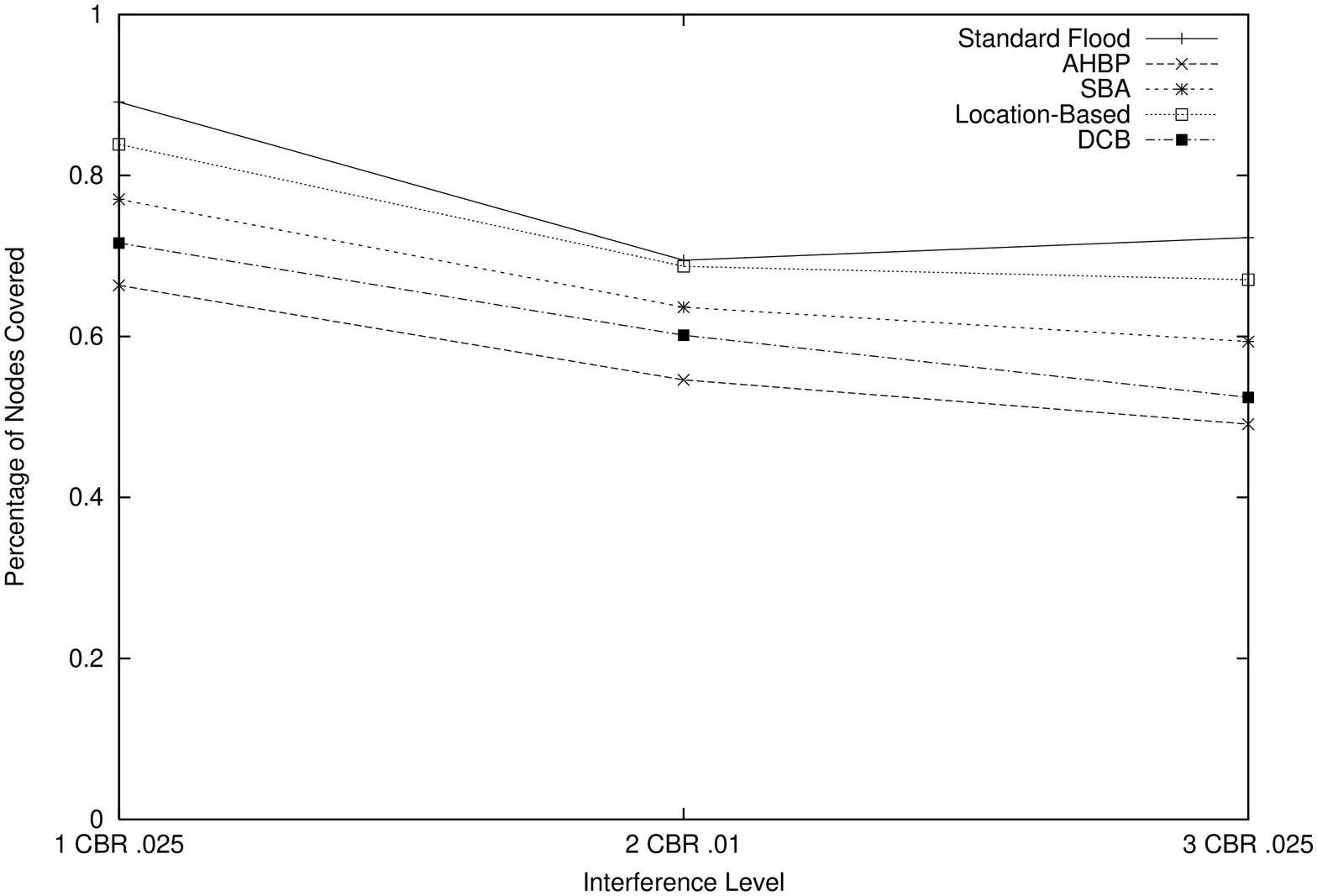,angle=270,width=0.95\linewidth,silent=}}
\caption{{\small Coverage, NWB Protocols, CBR Interference}}\label{prob4}
\end{minipage}
\begin{minipage}{3.5in}
\centerline{\epsfig{file=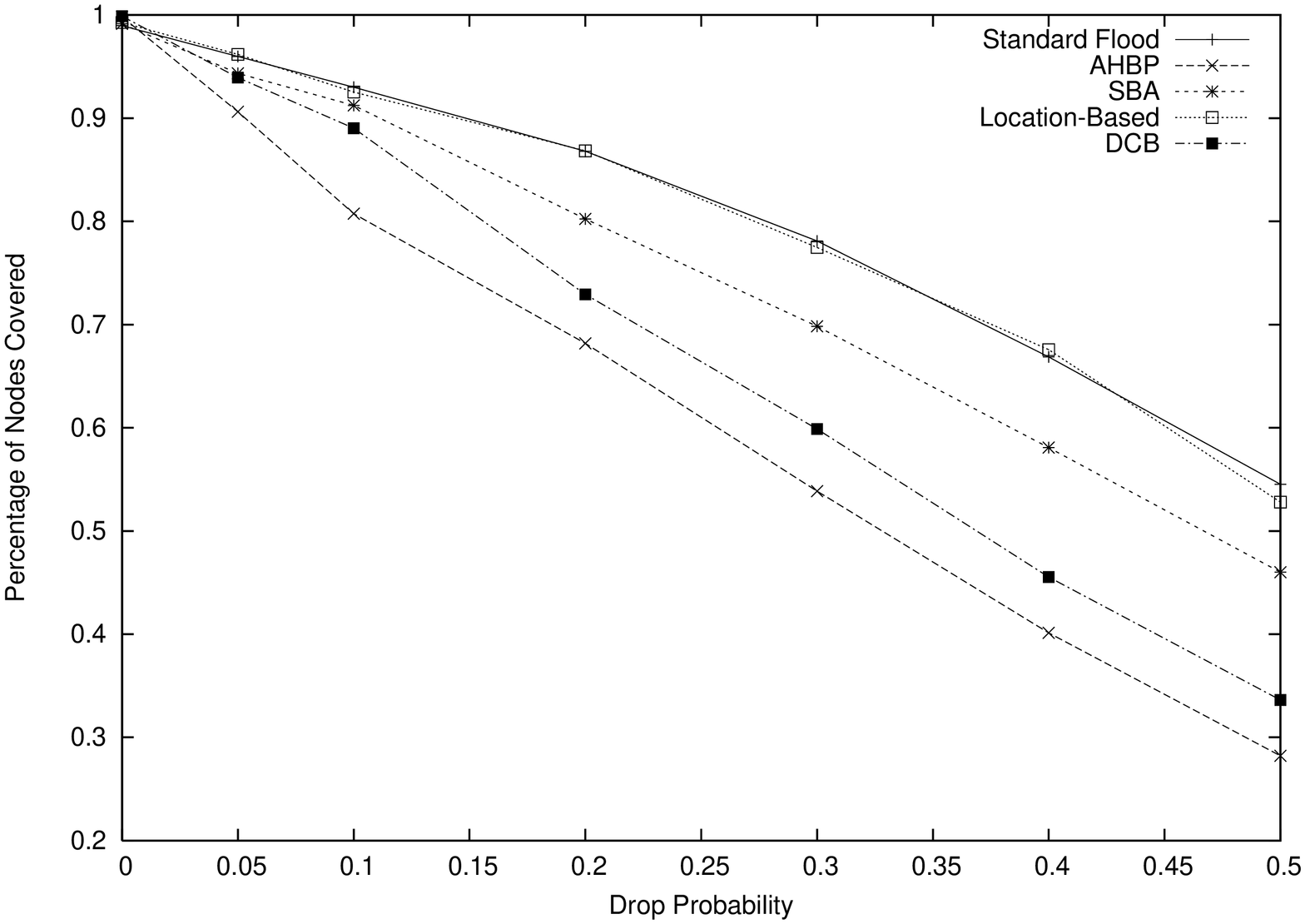,angle=270,width=0.95\linewidth,silent=}}
\caption{{\small Node Coverage, Controlled Drop}}\label{prob5}
\end{minipage}
\end{figure}

Figure~\ref{prob4} and Figure~\ref{prob5} show the coverage obtained
by the different NWB alternatives with CBR interference and
probabilistic dropping respectively for scenarios with 30 nodes.
Several observations can be made on this figure: (1) coverage degrades
for all 5 approaches as the interference increases.  This fact that
this occurs is not surprising, since all these protocols rely on the
unreliable MAC level broadcast.  However, the severity of the problem
is surprising ; (2) flooding remains the most reliable approach
because the other approaches reduce the redundancy available in
flooding; (3) the static algorithms (AHBP and DCB) perform worse than
the dynamic approaches because the forwarding responsibilities are
determined statically making it difficult to recover from losses.
Contrast this with the location based approach where other nodes may
rebroadcast the packet if a broadcast is lost since their local
coverage threshold will not be exceeded.  The poor performance is
somewhat surprising in the case of DCB because of its double-coverage
feature.  

\begin{figure}[ht]
\begin{minipage}{3.5in}
\centerline{\epsfig{file=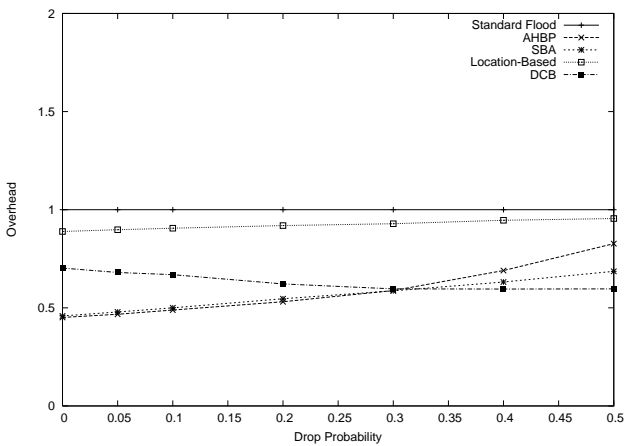,angle=270,width=0.95\linewidth,silent=}}
\caption{{\small Normalized Overhead}}\label{prob6}
\end{minipage}
\end{figure}

Figure~\ref{prob6} shows the normalized overhead of the different NWB
approaches.  As expected, AHBP, DCB, and SBA have a much lower
overhead than flooding, due to their neighborhood knowledge
algorithms.  However, note that for topology sensitive algorithms
(AHBP, DCB and SBA), this graph does not factor in the cost of the
'HELLO' messages that are periodically sent to accomplish neighborhood
discovery.  Neighbor discovery cost is more expensive than the cost of
flooding since every node sends an update and receives several
updates; it is even more expensive if two hop neighbors are tracked as
is the case in SBA.  However, this cost may be amortized over several
NWBs if multiple NWBs take advantage of a single round of discoveries.
Thus, the overhead shown in Figure~\ref{prob6} is a lower bound for
the topology-sensitive protocols, that can be quite optimistic if
neighborhood discovery frequency is not much lower than NWB frequency.

  The overhead of AHBP and SBA increases with the loss probability.
We conjecture that this occurs because in these low redundancy
algorithms, many nodes are covered by exactly one CDS member.  As the
probability of the loss of such broadcasts increases, the effective
coverage drops per transmission, leading to higher normalized
overhead.  The higher redundancy in the other algorithms mutes this
effect -- the loss of a transmission does not necessarily lead to
coverage loss if other redundant transmissions cover the same area.

The normalized overhead works well for comparing the relative overhead
of the different NWB algorithms.  However, it does not provide a
measure for the absolute overhead in terms of the stress it places on
the network, consuming power and bandwidth.  This absolute overhead is
application sensitive.  For example, if NWBs are used for discovering
paths in routing protocols, a single NWB may uncover multiple paths
which can be used to transport thousands of packets before another NWB
is needed.  In such an application, the overhead of the NWB is
marginal compared to the total number of packets sent; increasing the
overhead to improve the reliability of the flood does not detract from
the performance of the network appreciably.  Alternatively, in an
application where NWBs are dominant, the efficiency of the NWB is more
important.

\begin{figure}[ht]
\begin{minipage}{3.5in}
\centerline{\epsfig{file=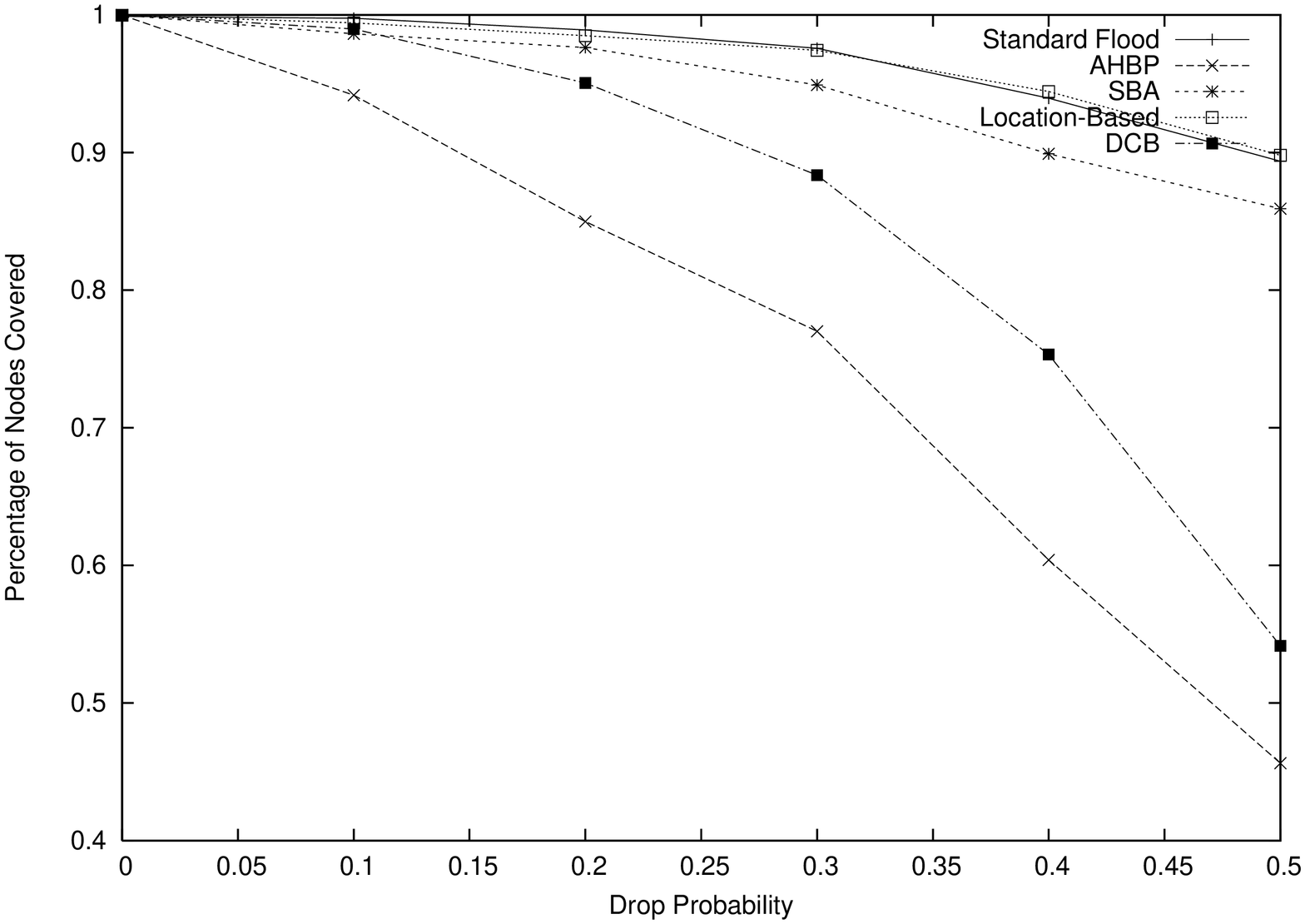,angle=270,width=0.95\linewidth,silent=}}
\caption{{\small Coverage, 50 node scenario, Controlled Drop}}\label{prob-50-cover}
\end{minipage}
\begin{minipage}{3.5in}
\centerline{\epsfig{file=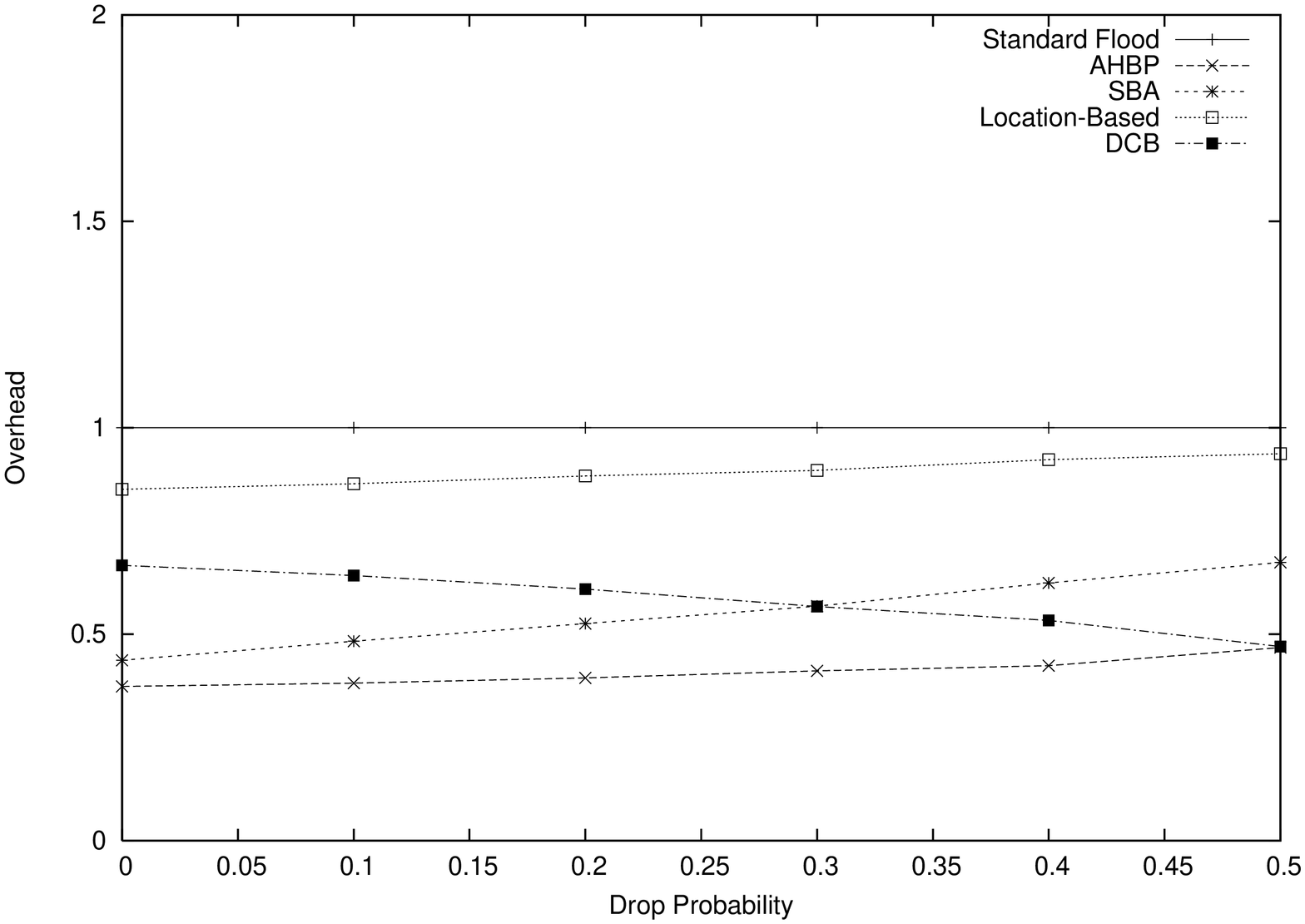,angle=270,width=0.95\linewidth,silent=}}
\caption{{\small Overhead, 50 node scenario, Controlled
Drop}}\label{prob-50-overhead}
\end{minipage}
\end{figure}

With increasing density, the performance of the NWB protocols improve,
but the problem remains.  This is demonstrated in
Figure~\ref{prob-50-cover} and Figure~\ref{prob-50-overhead} which
show the coverage and overhead for the case of 50 nodes.  Again, the
the static approaches (AHBP and DCB) perform badly in terms of coverage, but 
have the lowest overhead.  The dynamic
approaches perform much better than they do in the 30 node case as
they benefit from the increased available redundancy.  We focus on the
30 node scenario in our experiments because sparse areas of a network
are the most vulnerable/challenging to the flood unreliability
problem.  In addition, we present sample results with denser
scenarios.

\section{Proposed Solutions}~\label{solutions}
Increasing the flood reliability requires increasing the probability
of the reception of flood rebroadcast operations.  This is especially
true {\em in situations where their loss is likely or the redundancy
in the network is low} (e.g., under high interference or low
connectivity).  We seek to improve the reliability of the flood,
rather than insure complete reliability; guaranteeing reliability
requires too much overhead (for example, neighbor discovery and the
use of unicast packets).  Two classes of solutions can be identified:
(1) MAC level; and (2) Network level.

\subsection{MAC Level Solutions}
This class of solutions modifies the operation of the MAC protocol to
increase reliability.  Tang and Gerla~\cite{tang-00} proposed a scheme
for increasing the reliability of MAC broadcast that works as follows.
They require that broadcasts be acknowledged.  If the broadcast is not
acknowledged by at least one receiver, they retransmit the packet.
Requiring that receiver acknowledgment may result in a number of
receiving nodes concurrently sending an acknowledgment, leading to a
collision at the receiver.  They address this possibility by assuming
that any noise heard on the channel when the acknowledgment is
expected is due to a collision on the acknowledgment (no
retransmission is needed).  It is not clear how accurate this
assumption is in the presence of other traffic in the network.
Moreover, while this approach may guarantee that at least one node has
received the broadcast, it does so at the cost of generating an
acknowledgment from all receiving neighbors (an overhead that can
exacerbate unreliability if the network is already under high load).

We identify an alternative MAC layer solution which we call {\em
directed broadcast}.  In this approach, one of the neighbors is tasked
with responsibility for handshaking on the broadcast.  The broadcast
packet specifies the identity of the responsible neighbor.  All nodes
that receive the packet treat it as a broadcast packet; only the
identified neighbor acknowledges the packet.  It is also possible to
carry out full RTS/CTS/Broadcast/ACK dialog with this neighbor but
this is unlikely to be useful since broadcast packets tend to be
short.  The choice of the neighbor can be simple (any neighbor known
from a currently active connection, a recent flood, or through
listening promiscuously on nearby transmissions), or sophisticated
(e.g., proactively discovering neighbors and choosing the one most
likely to benefit the flood). Directed broadcast would be ideal for
protocols that require neighborhood knowledge such as
SBA~\cite{peng-00} and AHBP~\cite{peng-02}.  While it guarantees
delivery of the broadcast only to one of the neighbors, this increases
the overall success of the MAC broadcast~\cite{tang-00}.

Changes to the MAC protocol require redesign of the wireless cards and
the MAC standard and therefore face a high deployment barrier.
Moreover, they cannot easily adapt to the surrounding conditions,
which may lead to unnecessary overhead if the network is sparse or
lightly loaded.  For this reason we do not pursue these solutions
further in this paper, and instead focus on network layer solutions.

\subsection{Network Layer Solutions}
Lou and Wei discussed a network layer solution to increasing CDS-based
NWB algorithms~\cite{lou-04}; this approach was discussed in detail in
Section~\ref{related}.  The approach we pursue in this paper is
Selective Rebroadcast (SR): NWB packets are selectively rebroadcast if
they are suspected to have been lost.  We trade-off an increase in
overhead for an increase in reliability/node coverage.  This solution
is dual to the solutions that attempt to reduce the broadcast storm
problem~\cite{ni-99} by cutting down on the redundancy.  Those
solutions selectively eliminate rebroadcasts of the flood packet if it
is suspected to be redundant to cut down on the flood overhead.  In
contrast, we propose rebroadcasting important packets an additional
time if it is suspected to be lost in order to increase reliability.

It is important to note that SR should be applied judiciously.  In
dense regions of the network topology, the available redundancy is
high and the broadcast algorithm should cut down on the number of
rebroadcasts, especially if interference is low.  However, in sparse
regions of the network topology, and in the presence of interference,
SR can significantly improve the reliability of the NWB.

We explore the following criteria to decide when to rebroadcast a packet
an additional time.  This criteria mirrors ones proposed as solutions
for the broadcast storm problem~\cite{ni-99}.  However, while they
determine an upper threshold beyond which a rebroadcast is canceled,
we use a lower threshold beyond which a rebroadcast is repeated.
\begin{enumerate}
\item Probabilistic Solution: a packet is rebroadcast with a fixed
probability. In general, this solution is problematic because it does
not adapt to the density or loss rates in the network.
Therefore, it may result in large increases in the overhead when it is
not needed (e.g., in dense areas and/or when interference is low).

\item Counter-Based Solution: if the node does not hear \emph{n} other
nodes rebroadcast the packet within a certain amount of time, it will
rebroadcast it again.  This solution is attractive because it
naturally adapts to the interference level and density of the
network.  In a dense/low interference area, a number of rebroadcasts
is likely to be received after a node rebroadcasts a packet.  But in
sparse/high interference areas, this is not the case, and the
algorithm rebroadcasts a packet to enhance reliability.
\end{enumerate}

\noindent
In this work, we do not explore other criteria such as measuring the
interference level or the density of the network (e.g., using hello
messages) to determine when to rebroadcast the packet even though
density information is available in some of the NWB algorithms we
explore (e.g., SBA and AHBP).  For example, the probabilistic approach
can be biased with the current interference level (e.g., measured as
the observed utilization of the channel at the MAC
level~\cite{ahn-02}) such that the probability of rebroadcast is
increased if interference is high.  Both such schemes and MAC level
approaches are in our future work plans.

\section{Experimental Evaluation}~\label{experimental}
We implemented Selective Rebroadcast (SR) according to the criteria
presented in the previous section (probabilistic and counter based).
Since this criteria is computed based on locally available
information, SR can be added to all of the NWB algorithms we are
considering as follows.  For the probabilistic approach, a rebroadcast
timer is set probabilistically after the initial rebroadcast.  In the
counter based approach, the rebroadcast is initially set with a
timeout value, and then canceled if $n$ other rebroadcasts are
overhead.  For the NWB algorithms other than flooding, the additional
rebroadcast is only scheduled if an initial rebroadcast is decided
upon by the algorithm.  For example, in AHBP where a subset of the
nodes is selected to relay the packet further, only those nodes that
have an original forwarding responsibility are tasked with an
additional rebroadcast according to the probabilistic or counter-based
criteria.  Unless otherwise stated, we use scenarios with 30 nodes in
a 1000 by 1000 meter area.

\subsection{Analysis of Flooding}
\begin{figure}[ht]
\begin{minipage}{3.5in}
\centerline{\epsfig{file=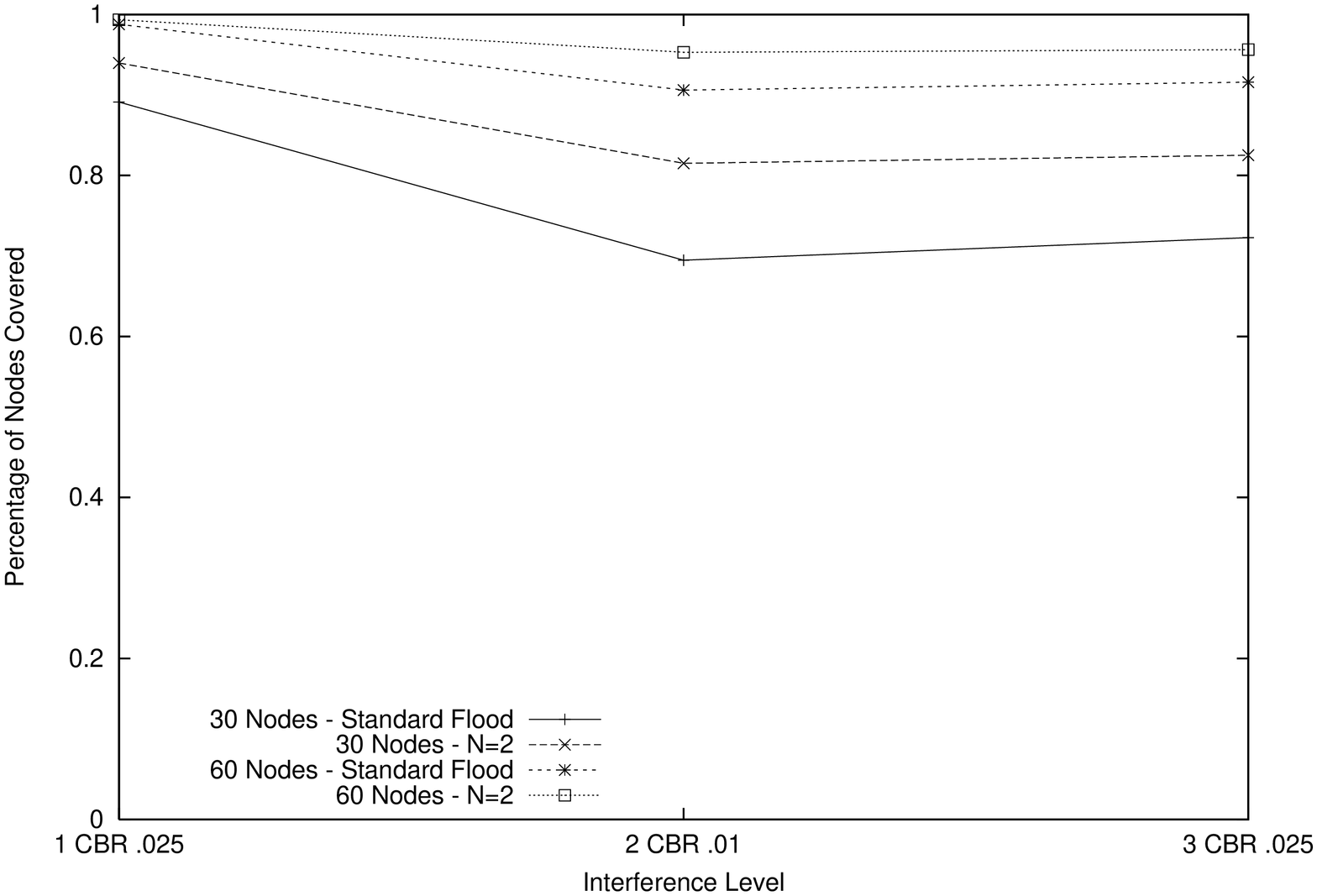,angle=270,width=0.95\linewidth,silent=}}
\caption{{\small Flooding with Counter SR (CBR Interference)}}\label{sol1-counter}
\end{minipage}
\begin{minipage}{3.5in}
\centerline{\epsfig{file=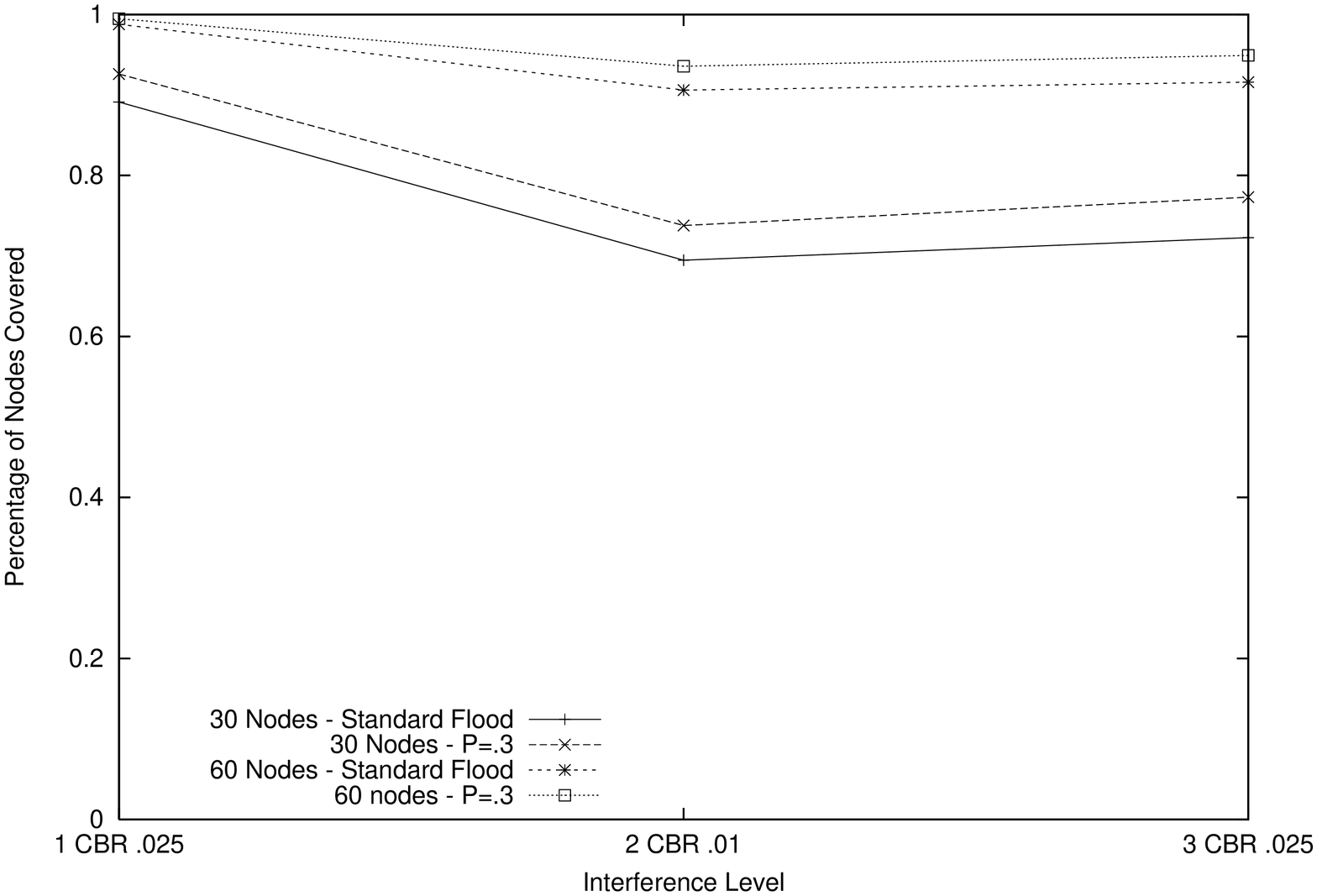,angle=270,width=0.95\linewidth,silent=}}
\caption{{\small Flooding with Probabilistic SR (CBR
Interference)}}\label{sol1-prob}
\end{minipage}
\end{figure}

Figure~\ref{sol1-counter} shows the effect of counter based
SR on the coverage of flooding under CBR interference.  Each
graph shows the coverage with standard flooding, and the improved
flooding for two density levels.  Figure~\ref{sol1-prob} shows the
coverage for the same scenarios with probabilistic rebroadcast.  While
probabilistic rebroadcast shows improvement in coverage, the
counter-based approach performs better because it is sensitive
to the density and interference levels.  

\begin{figure}[ht]
\begin{minipage}{3.5in}
\centerline{\epsfig{file=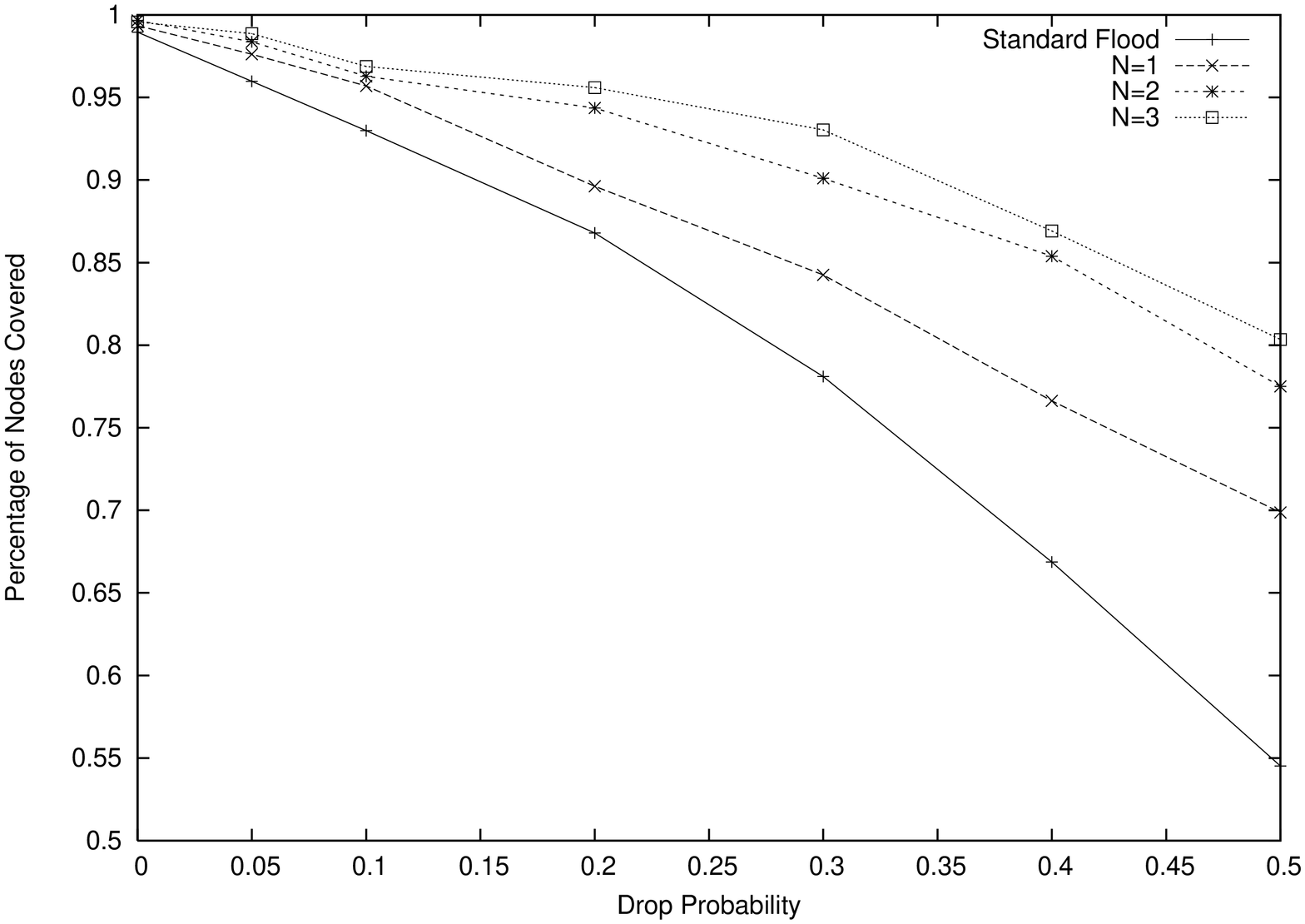,angle=270,width=0.95\linewidth,silent=}}
\caption{{\small Flooding with Counter SR (Controlled Drop)}}\label{sol2-counter}
\end{minipage}
\begin{minipage}{3.5in}
\centerline{\epsfig{file=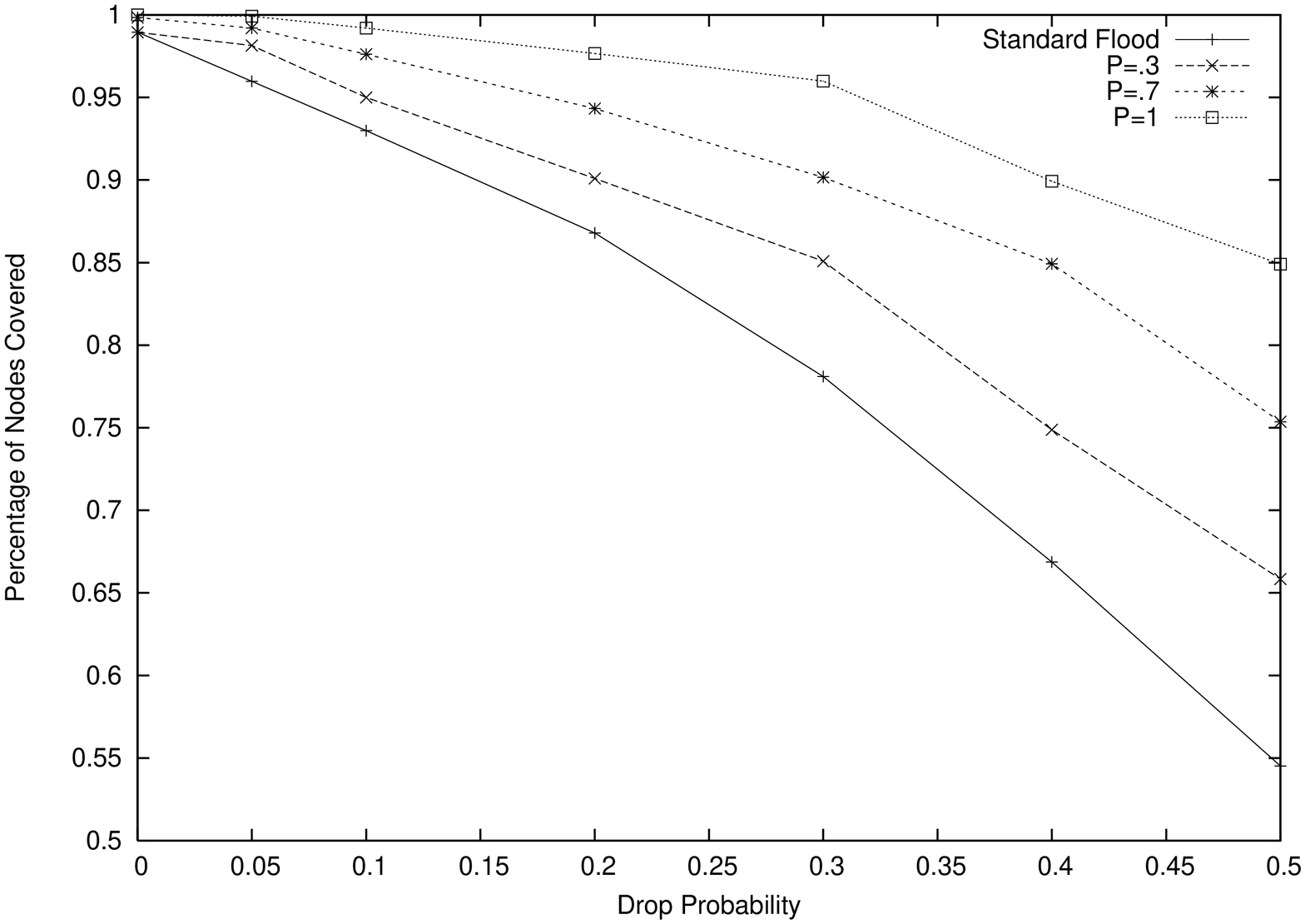,angle=270,width=0.95\linewidth,silent=}}
\caption{{\small Flooding with Probabilistic SR (Controlled Drop)}}\label{sol2-prob}
\end{minipage}
\end{figure}

The level of interference generated by the CBR connections is
unpredictable and non-uniform (interference will be generated around
the hops that make up the interfering connections).  Therefore, we
repeated the experiment with the controlled drop scenario, where
packets are dropped with a predetermined probability to simulate
interference (Figure~\ref{sol2-counter} and~\ref{sol2-prob}).  The
trend holds more clearly here with the improved solutions resulting in
a large improvement of coverage.  Again, counter-based solutions
outperform probabilistic rebroadcasts; for the remainder, we present results only with Counter SR.  

\begin{figure}[ht]
\begin{minipage}{3.5in}
\centerline{\epsfig{file=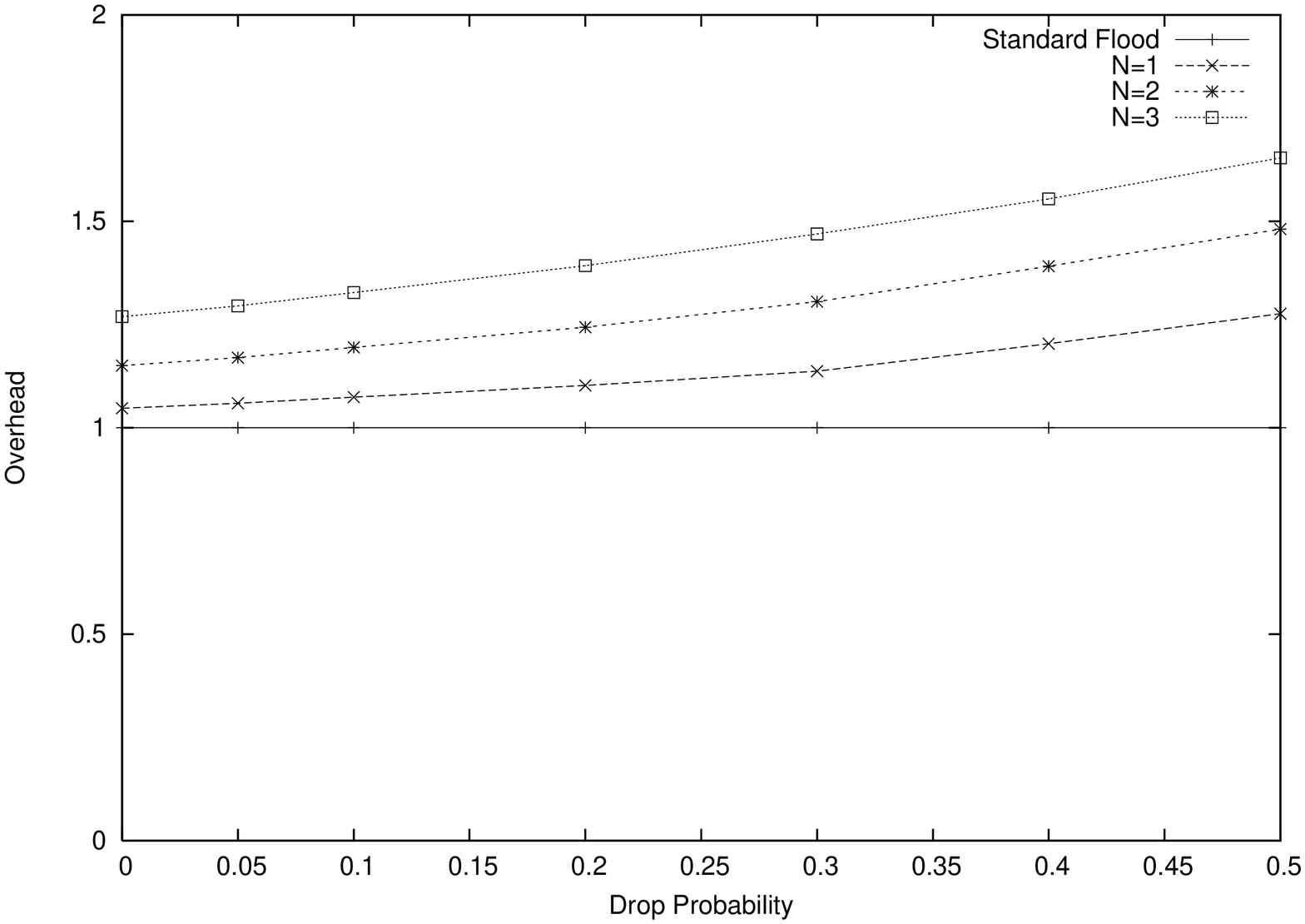,angle=270,width=0.95\linewidth,silent=}}
\caption{{\small Overhead (Flooding/Counter SR)}}\label{overhead-counter}
\end{minipage}
\begin{minipage}{3.5in}
\centerline{\epsfig{file=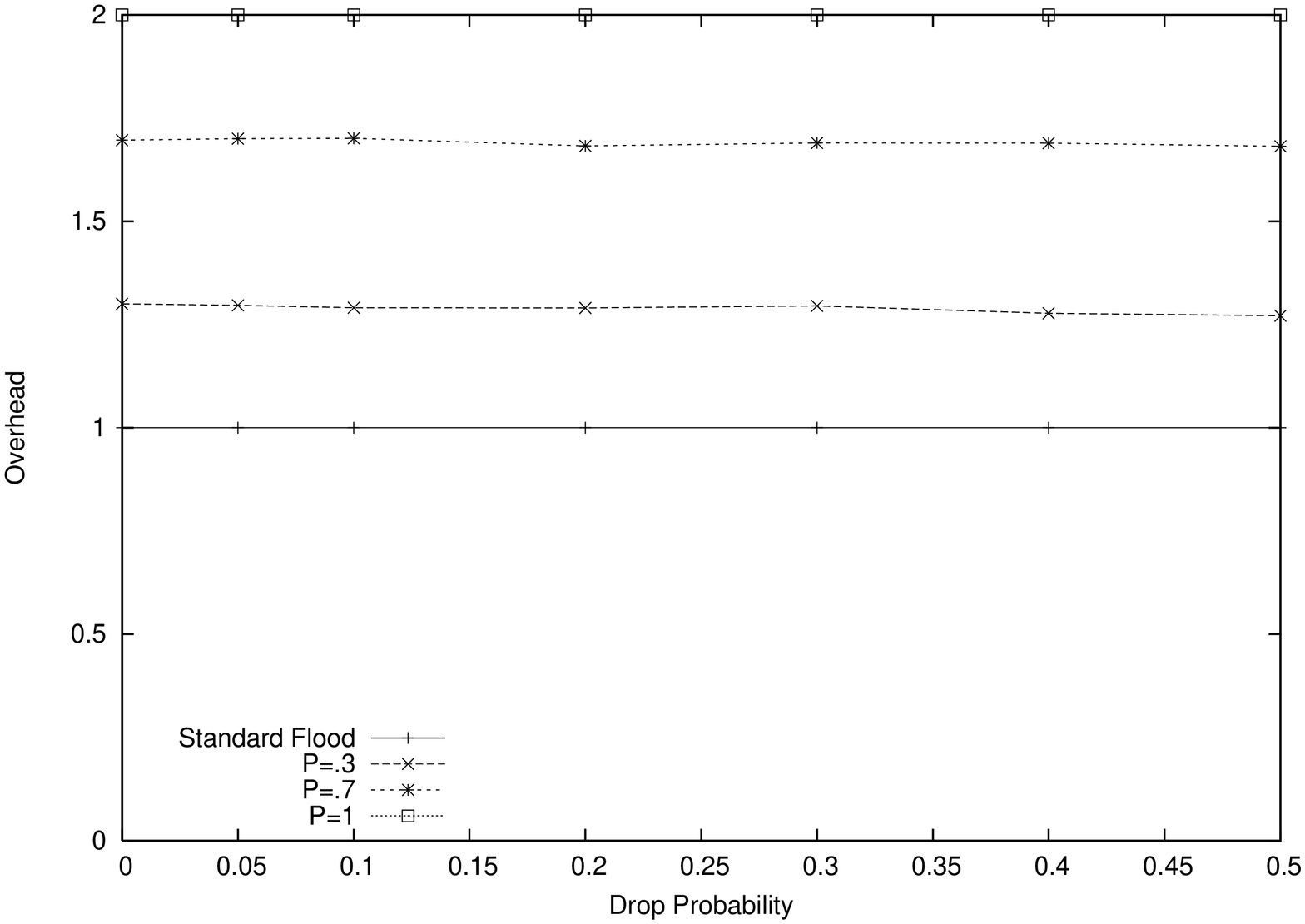,angle=270,width=0.95\linewidth,silent=}}
\caption{{\small Overhead (Flooding/Probabilistic SR)}}\label{overhead-prob}
\end{minipage}
\end{figure}

Figure~\ref{overhead-counter} and~\ref{overhead-prob} show the
normalized overhead of the flood augmented with SR for a controlled
drop scenario of 30 nodes.  The probabilistic approach results in an
increase in normalized overhead proportional to the rebroadcast
probability.  The counter based approach also results in an increase
in the flood overhead.  It is interesting to note how the overhead
increases with the drop probability indicating a key property of this
approach: as the loss rate increases, additional nodes compensate
dynamically and generate a rebroadcast.  Thus, the SR algorithm
automatically increases its coverage redundancy as the loss rate
increases.  The overhead of DCB does not follow the same pattern
because the degree of redundancy is statically fixed (at 2) regardless
of the loss rate experienced in the network.

\subsection{Other NWB Algorithms}
\begin{figure}[ht]
\begin{minipage}{3.5in}
\centerline{\epsfig{file=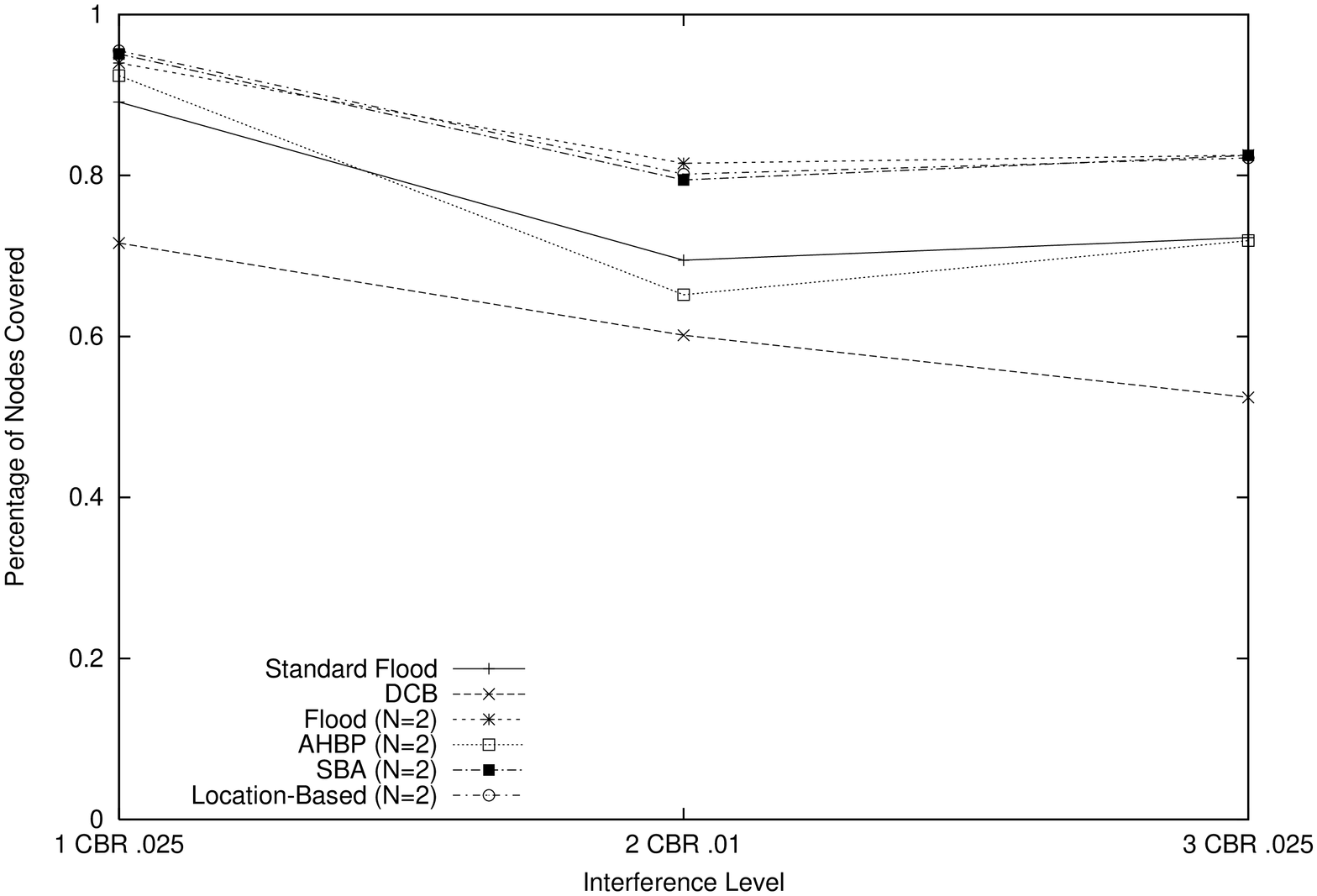,angle=270,width=0.95\linewidth,silent=}}
\caption{{\small Coverage (Counter SR)}}\label{mines-counter}
\end{minipage}
\begin{minipage}{3.5in}
\centerline{\epsfig{file=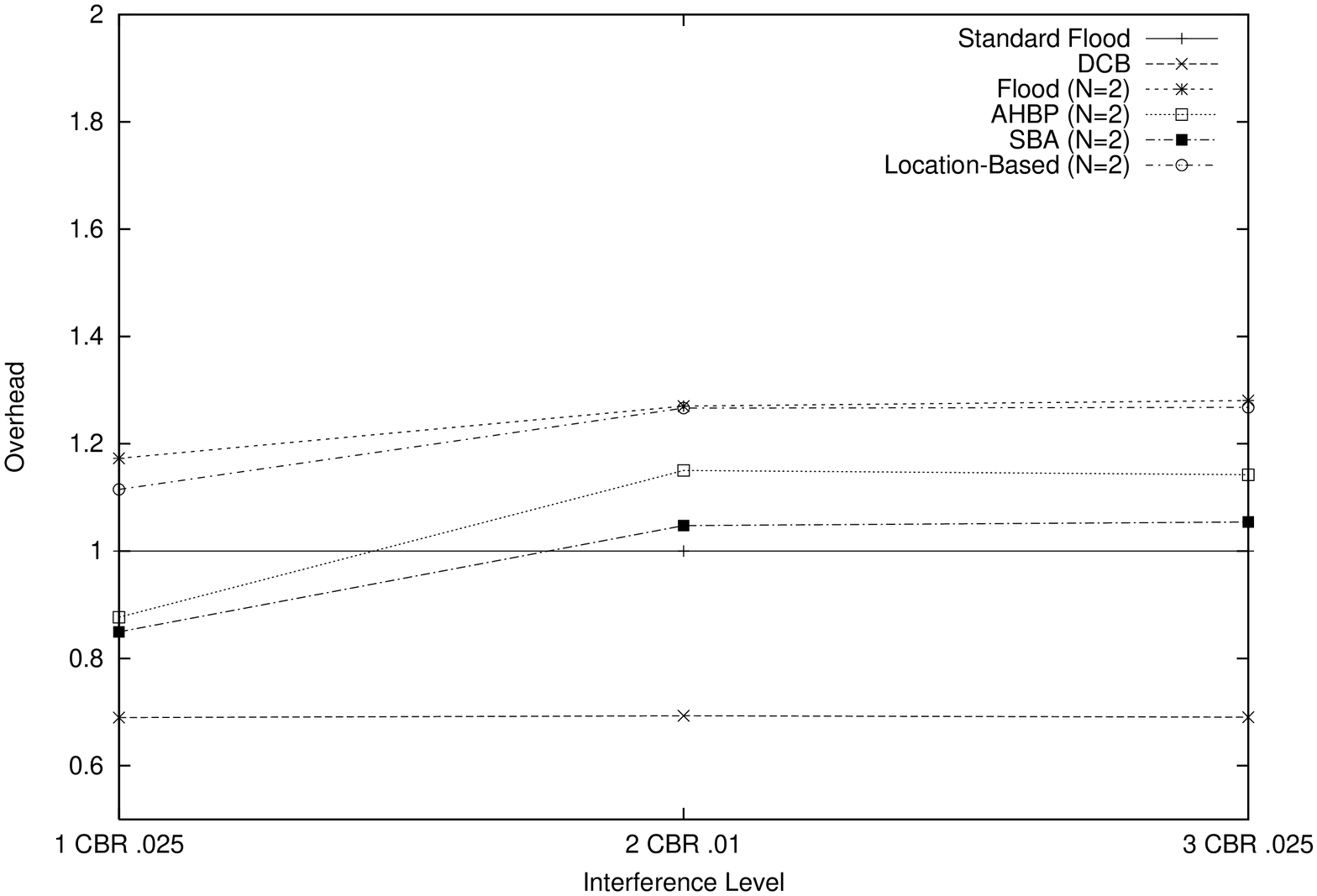,angle=270,width=0.95\linewidth,silent=}}
\caption{{\small Overhead (Counter SR)}}\label{counter-overhead}
\end{minipage}
\end{figure}

Figure~\ref{mines-counter} shows the effect of the counter-based SR on
the coverage of the other NWB algorithms with CBR interference.  In
all cases, Similarly, Figure~\ref{counter-overhead} shows the
normalized overhead with the counter based rebroadcast.  Recall that
the performance of the standard version of these algorithms performed
worse than flooding in terms of coverage (Figure~\ref{prob4}).  In all
cases, the coverage is improved well beyond flooding.  This additional
coverage comes at an increase in the overhead due to the additional
rebroadcasts as can be seen in Figure~\ref{counter-overhead}.  We
believe that the rather high overhead for the counter based approach
is skewed due to the nodes at the edge of the network.  Such nodes do
not have many neighbors and end up not receiving enough rebroadcasts
to satisfy the counter threshold, leading to unnecessary rebroadcasts.
Utilizing neighborhood knowledge would result in appreciable
improvement of this problem by adapting the threshold to the number of
neighbors.  This is a topic of our future research.

LBA shows a modest increase in overhead over standard flooding, but a
large increase in reliability.  This is due to the fact that it adapts
its retransmission to the density and level of interference in the
network.  This is also true for SBA (and partly true for AHBP).  The
overhead for these protocols is lower than flooding; however, for SBA
and AHBP, the overhead figure does not include the ongoing overhead of
maintaining neighborhood knowledge.  Finally, we believe that the
overhead results are somewhat inflated due to boundary nodes where
additional rebroadcasts would be unecessarily generated.  This effect
can be reduced or eliminated with neighborhood knowledge.

\begin{figure}[ht]
\begin{minipage}{3.5in}
\centerline{\epsfig{file=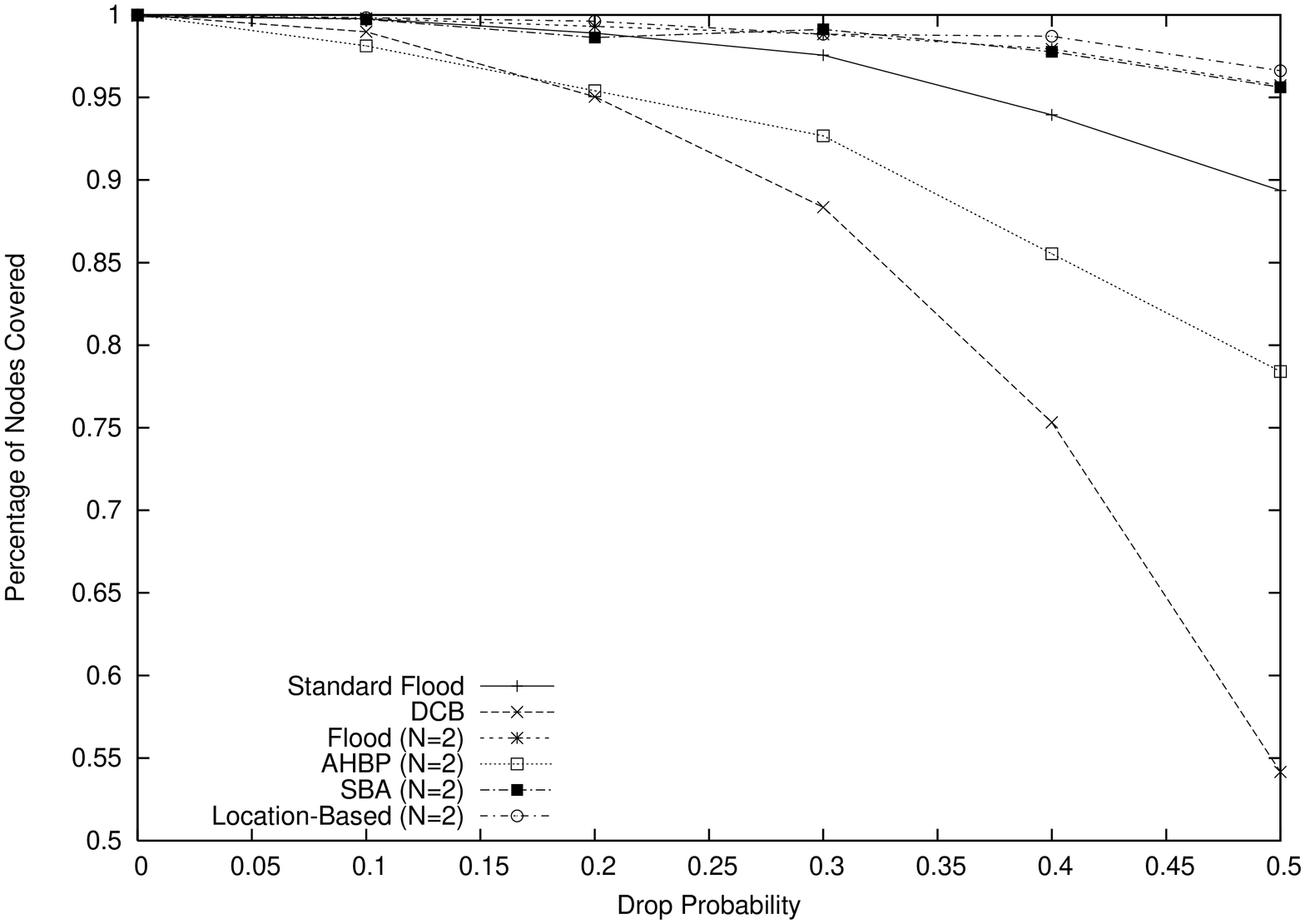,angle=270,width=0.95\linewidth,silent=}}
\caption{{\small Coverage (Counter SR), 50 Nodes}}\label{mines-counter-50}
\end{minipage}
\begin{minipage}{3.5in}
\centerline{\epsfig{file=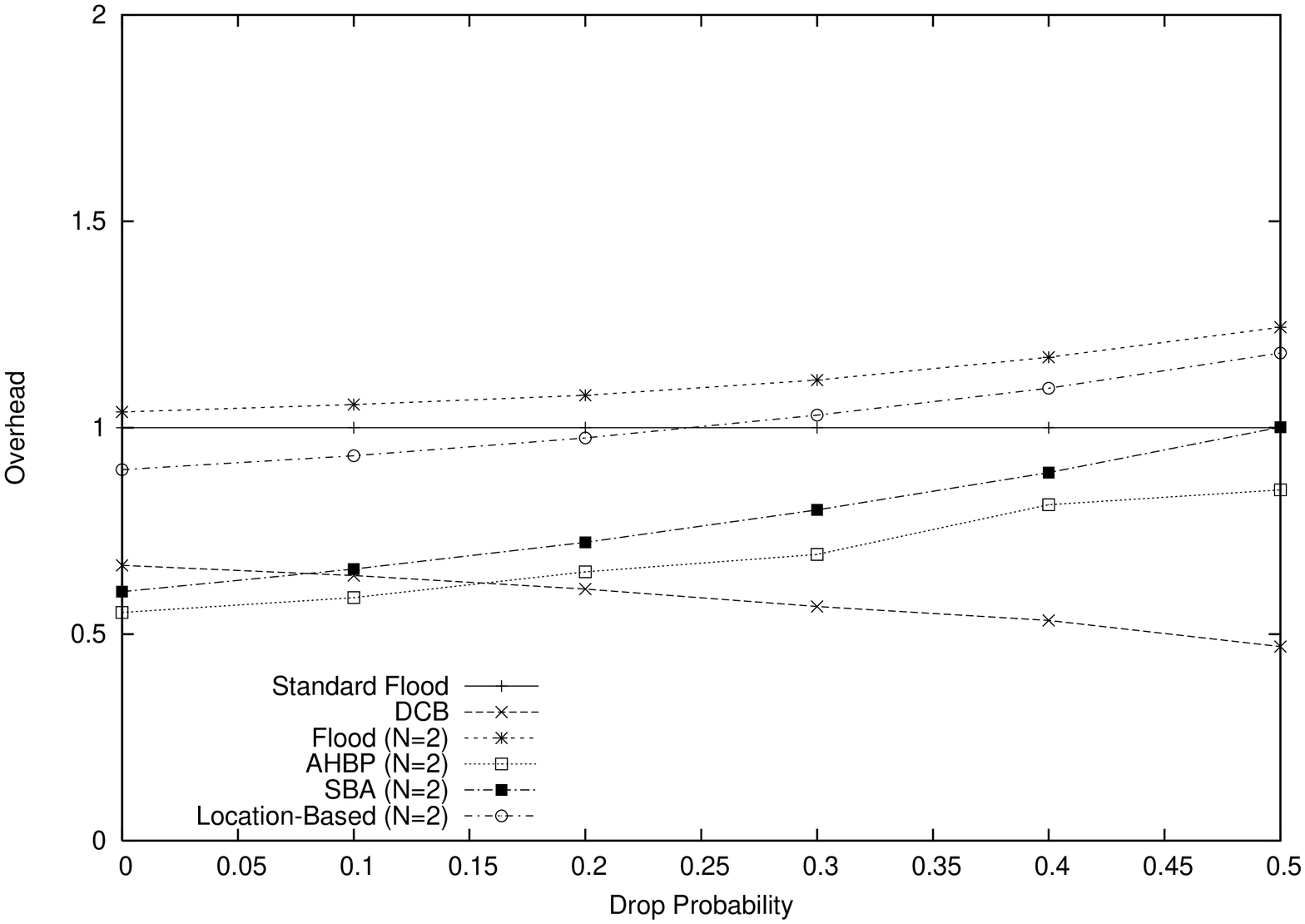,angle=270,width=0.95\linewidth,silent=}}
\caption{{\small Overhead (Counter), 50 Nodes}}\label{counter-overhead-50}
\end{minipage}
\end{figure}

Figure~\ref{mines-counter-50} presents the coverage achieved by SR in
the more dense 50 node scenario.  These results are shown to
demonstrate that solution applies at higher density.  Again, coverage
is considerably improved with SR at a modest cost in terms of
overhead.  We believe that for many applications, such as route
discovery, the increased reliability is worth a small increase in
overhead.

\subsection{Effect of Mobility}
\begin{figure}[ht]
\begin{minipage}{3.5in}
\centerline{\epsfig{file=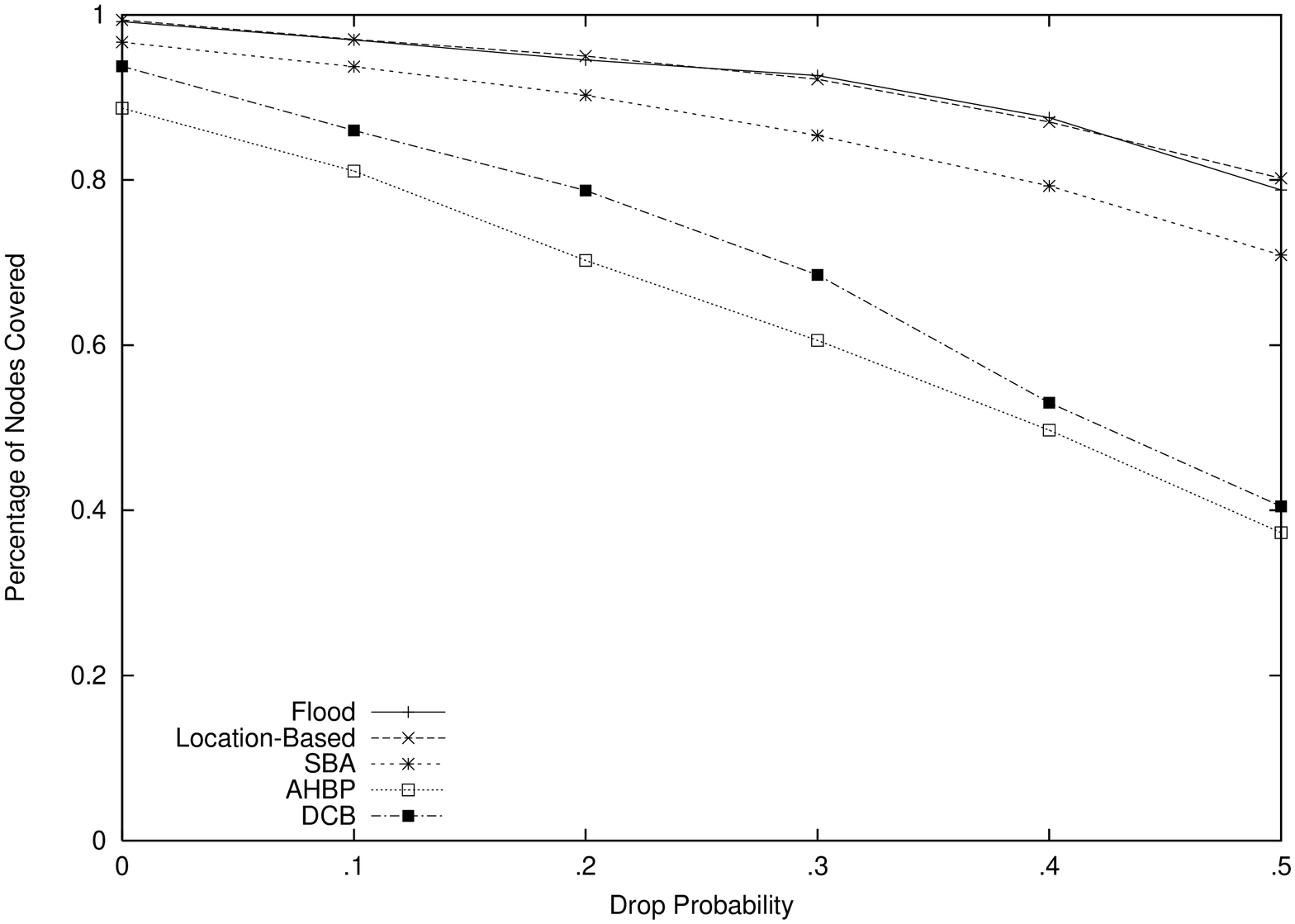,angle=270,width=0.95\linewidth,silent=}}
\caption{{\small Coverage, Mobile Scenarios (5 m/s)}}\label{mob1}
\end{minipage}
\begin{minipage}{3.5in}
\centerline{\epsfig{file=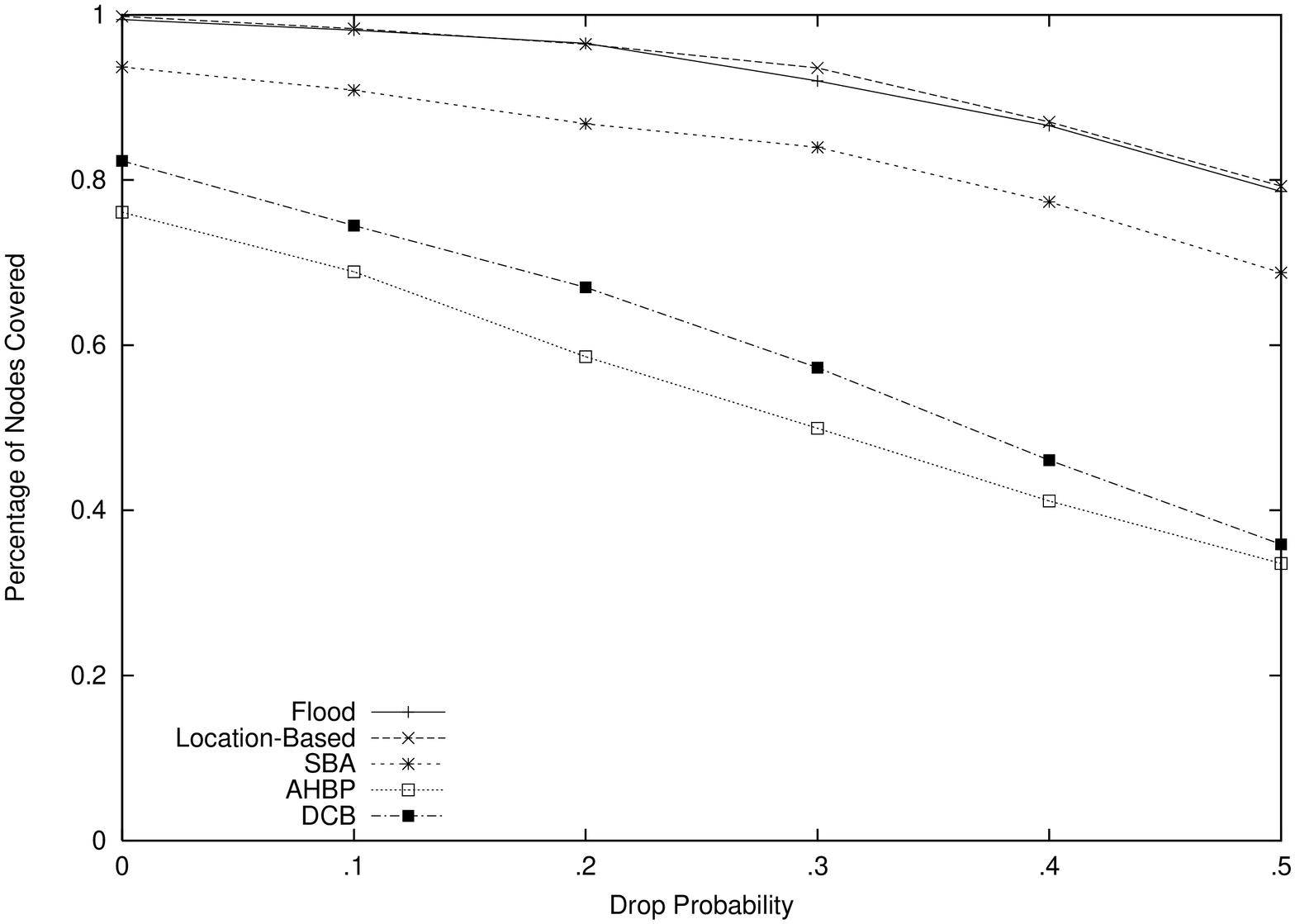,angle=270,width=0.95\linewidth,silent=}}
\caption{{\small Coverage, Mobile Scenarios (15 m/s)}}\label{mob2}
\end{minipage}
\end{figure}

\begin{figure}[ht]
\begin{minipage}{3.5in}
\centerline{\epsfig{file=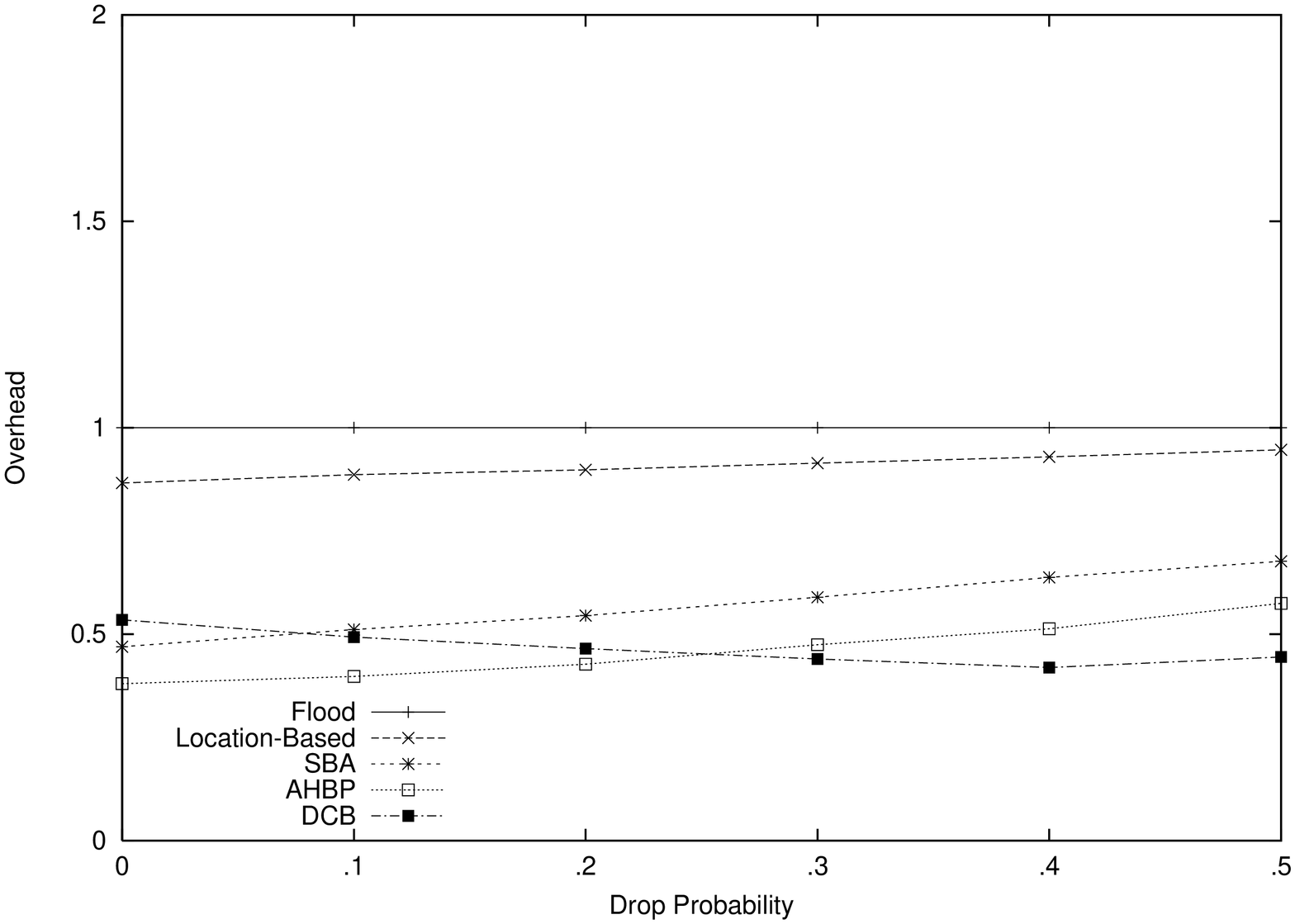,angle=270,width=0.95\linewidth,silent=}}
\caption{{\small Overhead, Mobile Scenarios (15 m/s)}}\label{mob2-ov}
\end{minipage}
\end{figure}

Figure~\ref{mob1} and Figure~\ref{mob2} show the effect of random
waypoint mobility with 5 m/s and 15 m/s average speed respectively
on node coverage achieved by the NWB algorithms.  Neighborhood
discovery was carried out every 1 second.  Even at such an agressive
neighbor discovery frequency, its clear that the performance of AHBP
and DCB degrades with mobility.  The dynamic approaches
appear more resilient to mobility.  Overhead did not appreciably
change with mobility, as can be seen in figure~\ref{mob2-ov} (again,
this overhead does not include the neighborhood discovery component).

\begin{figure}[ht]
\begin{minipage}{3.5in}
\centerline{\epsfig{file=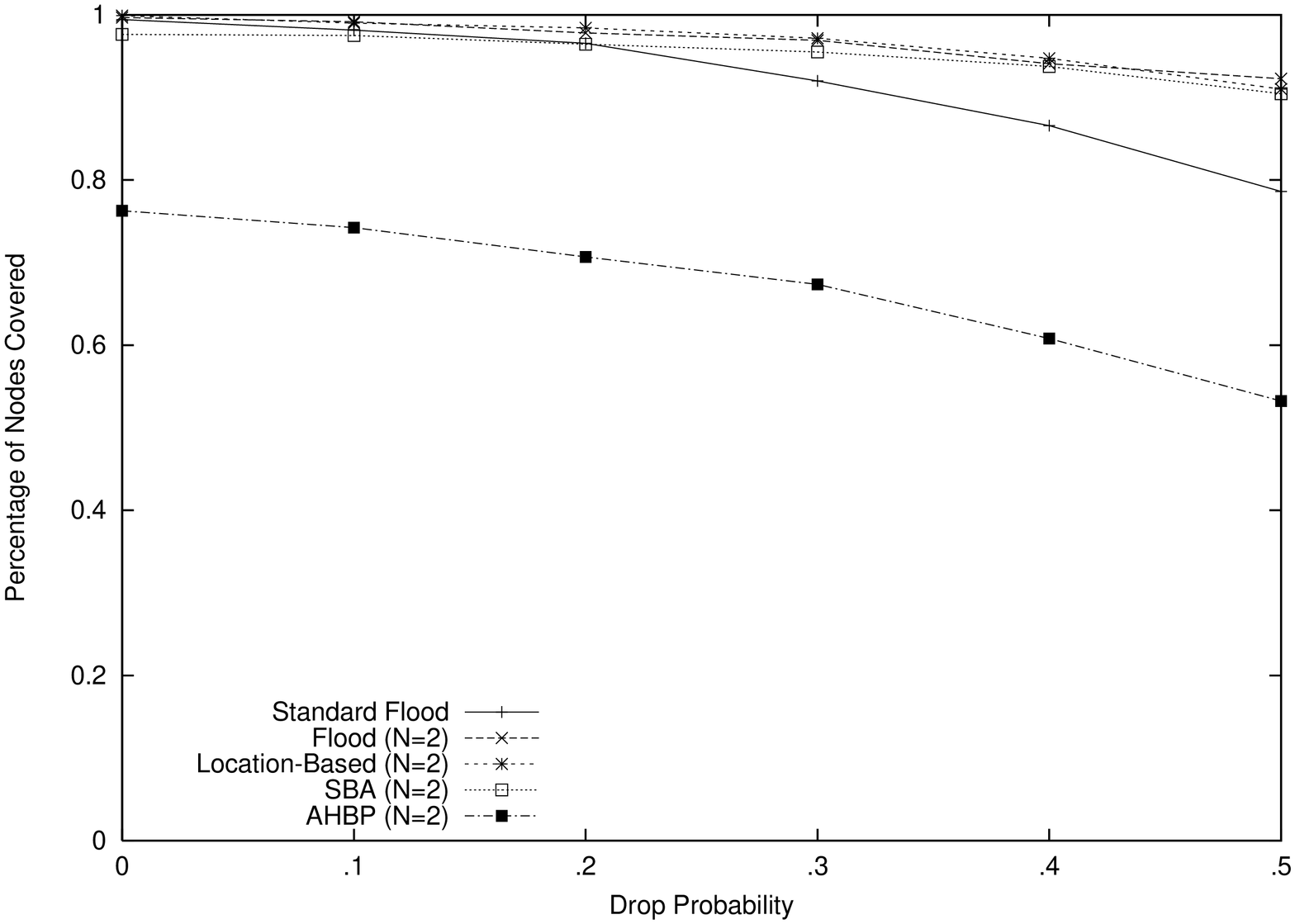,angle=270,width=0.95\linewidth,silent=}}
\caption{{\small Coverage with Counter SR and Mobility (15
m/s)}}\label{mob3}
\end{minipage}
\begin{minipage}{3.5in}
\centerline{\epsfig{file=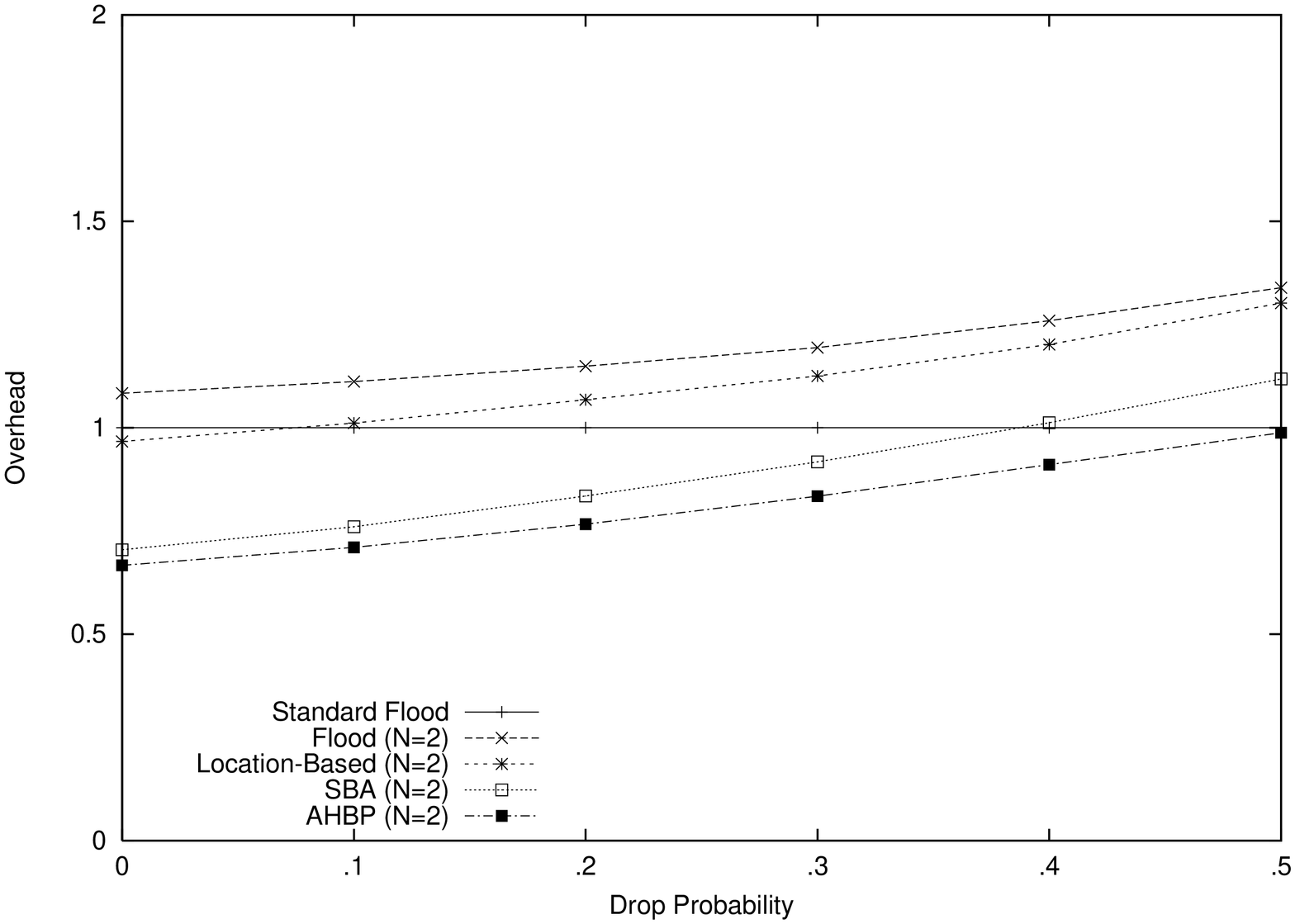,angle=270,width=0.95\linewidth,silent=}}
\caption{{\small Overhead with Counter SR and Mobility (15 m/s)}}\label{mob3-ov}
\end{minipage}
\end{figure}

Figure~\ref{mob3} shows the performance of the NWB algorithms with SR
(counter-based with 2 rebroadcast threshold).  Again, SR improved the
coverage of all protocols.  Overhead (Figure~\ref{mob3-ov}) did not
change perceptibly with mobility.

\section{Conclusions}~\label{conclude}
Network-Wide Broadcasts (NWBs) are important operations in Ad hoc
networks that are used in several routing and group communication
algorithms.  Existing research has targeted efficient NWB to reduce
the amount of redundancy inherent in flooding (the simplest NWB
approach).  As a result, the NWB becomes more susceptible to loss of
coverage due to transmission losses that result from heavy
interference or transmission errors.  This problem arises because NWBs
rely on an unreliable MAC level broadcast operation to reach multiple
nodes with one transmission for a more efficient coverage of the
nodes.  In the presence of interference or transmission errors, this
results nodes not receiving the NWB.  

We outline the NWB solution space and explain the properties of the
different solutions in terms of overhead, resilience to mobility as
well as reliability.  We demonstrate the problem using interference
from CBR connections as well as using probabilistic dropping of packets
(which allows us to control loss rate systematically).  As the
loss rate rises, and as the density of the network goes down,
the coverage achieved by an NWB operation
drops.  We showed that this is especially true
for other NWB algorithms that also rely on MAC level broadcasts
because they target reducing the redundancy present in floods, making
them more suceptible to losses.  Furthermore, we showed that
approaches that statically determine the set of forwarding nodes
perform much worse than ones that dynamically evaluate
whether their rebroadcast is likely to be redundant.  As a result, the
recent DCB algorithm, while it improves coverage relative to other
static approaches, remains substantially more vulnerable to losses
than dynamic approaches.

Several approaches to address this problem were discussed.  Improving
reliability at the MAC level requires a mechanism to allow
acknowledgments (or partial acknowledgment) of packet reception.
Tang and Gerla had proposed such an approach where receiving nodes
acknowledge a packet and acknowledgment collisions (detected as
noise) are interpreted as a successful broadcast.  We proposed another
approach (directed broadcast) with a much lower overhead and more
control on the receiving node.  

We did not evaluate MAC level approaches because of the deployment
barrier facing MAC level changes.  Instead, we focused on network
level solutions which can be directly implemented in the routing
algorithm.  The intuition behind our selective rebroadcast approach is
to selectively rebroadcast a packet if it is suspected that it has been
lost, especially if the network density is locally low.  We explored
two simple approaches: (1) probabilistic rebroadcast: a packet is
resent after the initial forwarding with a fixed probability; and (2)
counter-based rebroadcast: a packet is resent if we do not hear $n$
subsequent retransmissions of the same packet (indicating that other
nearby nodes have received it).

The two criteria were added to the four NWB algorithms (LBA, SBA, AHBP
and flooding).  The results show that a large improvement in coverage
results in all cases, with coverage higher than that of flooding, and
significantly higher than that of DCB.  However, this comes
at an increase in overhead that rises with the loss rate (to retain
coverage level in the face of losses).  LBA with selective rebroadcast
had lower overhead than flooding except at very high loss rates, with
a large improvement in coverage.  While the topology sensitive
algorithms (LBA, SBA, AHBP and DCB) have lower overhead, this does not
take into account the potentially high cost of neighborhood
discovery.

We believe that overhead can be reduced substantially with more
sophisticated policies for selective rebroadcast.  For example, the
nodes near the boundary of the simulated area for the counter-based
approach will almost always rebroadcast the packet unnecessarily
because they are likely to be leaf nodes for the broadcast and no
subsequent rebroadcasts will result; since the counter threshold is
not reached, this triggers unnecessary selective rebroadcasts.  We are
working on more effective policies that infer the local interference
and density and use this information to more intelligently assess when
selective rebroadcasts are useful.  We envision an NWB algorithm that
adaptively controls rebroadcasts to reduce redundancy while
maintaining reliability in the presence of interference.  This is one
of the areas we are exploring in our future research.

\bibliography{ad-hoc,mac,broadcast,local}

\begin{thebibliography}{10}

\bibitem{clausen-03}
T.~Clausen and P.~Jacquet,
\newblock ``Optimized link state routing protocol,''
\newblock {Internet Draft}, Internet Engineering Task Force, Oct. 2003,
\newblock http://www.ietf.org/internet-drafts/draft-ietf-manet-olsr-11.txt.

\bibitem{das-00}
S.~Das, C.~Perkins, and E.~Royer,
\newblock ``Performance comaprison of two on-demand routing protocols for ad
  hoc networks,''
\newblock in {\em Proc. of INFOCOM 2000}, Mar. 2000.

\bibitem{lewis-02}
M.~Lewis, F.~Templin, B.~Bellur, and R.~Ogier,
\newblock ``Topology broadcast based on reverse-path forwarding (tbrpf),''
\newblock {Internet Draft}, Internet Engineering Task Force, June 2002,
\newblock http://www.ietf.org/internet-drafts/draft-ietf-manet-tbrpf-06.txt.

\bibitem{royer-99a}
E.~Royer and C.~Perkins,
\newblock ``Multicast operation of the ad hoc on-demand distance vector routing
  protocol,''
\newblock in {\em Proc. of the ACM International Conference on Mobile Computing
  and Networking (MobiCom'99)}, 1999, pp. 207--218.

\bibitem{ni-99}
S.~Ni, Y.~Tseng, Y.~Chen, and J.~Sheu,
\newblock ``The broadcast storm problem in a mobile ad hoc network,''
\newblock in {\em Proceedings of ACM/IEEE International Conference of Mobile
  Computing and Networking (MOBICOM'99)}, Sept. 1999.

\bibitem{alzoubi-02}
K.~Alzoubi, {P.-J.} Wan, and O.~Frieder,
\newblock ``Message-optimal connected dominating sets in mobile ad hoc
  networks,''
\newblock in {\em Proceedings of the 3rd ACM international symposium on Mobile
  ad hoc networking and computing {(MobiHoc 2002)}}, 2002, pp. 157--164.

\bibitem{ghandi-03}
R.~Gandhi, S.~Parthasarathy, and A.~Mishra,
\newblock ``Minimizing broadcast latency and redundancy in ad hoc networks,''
\newblock in {\em Proceedings of the 4th ACM international symposium on Mobile
  ad hoc networking and computing {(MobiHoc 2003)}}, 2003, pp. 222--232.

\bibitem{decouto-02}
D.~{De Couto}, D.~Aguayo, B.~Chambers, and R.~Morris,
\newblock ``Performance of multihop wireless networks: Shortest path is not
  enough,''
\newblock in {\em Proceedings of the First Workshop on Hot Topics in Networks
  (HotNetsI)}, Oct. 2002.

\bibitem{lou-04}
W.~Lou and J.~Wu,
\newblock ``Double-covered broadcast (dcb): A simple reliable broadcast
  algorithm in manets,''
\newblock in {\em Proc. of INFOCOM 2004}, 2004.

\bibitem{pagani-97}
E.~Pagani and G.~Rossi,
\newblock ``Reliable broadcast in mobile multihop packet networks,''
\newblock in {\em Proceedings of ACM MOBICOM'97}, Sept. 1997, pp. 34--42.

\bibitem{lipman-04}
J.~Lipman, P.~Boustead, and J.~Chicharo,
\newblock ``Reliable optimised flooding in ad hoc networks,''
\newblock in {\em Proceedings of the IEEE 6th CAS Symposium on Emerging
  Technologies: Frontiers of Mobile and Wireless Communication}, May 2004.

\bibitem{bhagarvan-94}
V.~Bharghavan, A.~Demers, S.~Shenker, and L.~Zhang,
\newblock ``{MACAW}: A media access protocol for wireless lan's,''
\newblock in {\em Proc. SIGCOMM '94}, 1994, pp. 212--225.

\bibitem{fullmer-97}
C.~L. Fullmer and J.~J. {Garcia-Luna-Aceves},
\newblock ``Solutions to hidden terminal problems in wireless networks,''
\newblock in {\em Proceedings of SIGCOMM 1997}, 1997, pp. 39--49.

\bibitem{crow-97}
B.~Crow, I.~Widjaja, J.~Kim, and P.~Sakai,
\newblock ``{IEEE} 802.11 wireless local area networks,''
\newblock {\em IEEE Communications Magazine}, pp. 116--126, Sept. 1997.

\bibitem{sivakumar-03}
R.~Sivakumar, P.~Sinha, and V.~Bharghavan,
\newblock ``Braving the broadcast storm: Infrastructural support for ad hoc
  routing,''
\newblock {\em Computer Networks: The International Jorunal of Computer and
  Telecommunication Networking}, vol. 41, no. 6, pp. 687--706, Apr. 2003.

\bibitem{lim-00}
H.~Lim and C.~Kim,
\newblock ``Multicast tree construction and flooding in wireless adhoc
  networks,''
\newblock in {\em Proceedings of the ACM International Workshop on Modeling,
  Analysis and Simulation of Wireless and Mobile Swstems ({MSWIM})}, 2000.

\bibitem{peng-00}
W.~Peng and X.~Lu,
\newblock ``On the reduction of broadcast redundancy in mobile ad hoc
  networks,''
\newblock in {\em Proc. of MobiHoc 2000}, 2000.

\bibitem{tang-00}
K.~Tang and M.~Gerla,
\newblock ``{MAC} layer broadcast support in 802.11 networks,''
\newblock in {\em Proc. of IEEE MILCOM 2001}, Oct. 2001, pp. 544--548.

\bibitem{ns-2}
UC~Berkeley/LNBL/ISI,
\newblock ``The ns-2 network simulator with the cmu mobility extensions,''
  2002,
\newblock \url{http://www.isi.edu/nsnam/ns/}.

\bibitem{xu-02}
S.~Xu and T.s Saadawi,
\newblock ``Revealing the problems with 802.11 medium access control protocol
  in multi-hop wireless ad hoc networks,''
\newblock {\em Computer Networks}, vol. 38, no. 4, pp. 531--548, Mar. 2002.

\bibitem{peng-02}
W.~Peng and X.~Lu,
\newblock ``{AHBP}: An efficient broadcast protocol for mobile ad hoc
  networks,''
\newblock {\em Journal of Science and Technology (Beijing, China)}, 2002.

\bibitem{ahn-02}
{G.-S.} Ahn, {L.-H.} Sun, A.~Veres, and A.~Campbell,
\newblock ``{SWAN}: Service differentiation in stateless wireless ad hoc
  networks,''
\newblock in {\em Proc. of INFOCOM 2002}, 2002.

\end{thebibliography}
\bibliographystyle{IEEE}

\end{document}